\documentclass[12pt,preprint]{aastex}

\shorttitle{Yellow Supergiants in the SMC}

\shortauthors{Neugent et al.}

\begin{document}

\title{Yellow Supergiants in the Small Magellanic Cloud (SMC): \\ Putting Current Evolutionary Theory to the Test}

\author{Kathryn F. Neugent, Philip Massey\altaffilmark{1} and Brian Skiff}
\affil{Lowell Observatory, 1400 W Mars Hill Road, Flagstaff, AZ 86001; kneugent@lowell.edu; phil.massey@lowell.edu; bas@lowell.edu}

\author{Maria R. Drout\altaffilmark{1}}
\affil{Department of Physics \& Astronomy, University of Iowa, Iowa City, IA 52245; maria-drout@uiowa.edu}

\author{Georges Meynet}
\affil{Geneva University, Geneva Observatory, CH-1290 Versoix, Switzerland; georges.meynet@unige.ch}

\author{Knut A. G. Olsen}
\affil{National Optical Astronomy Observatory, 950 North Cherry Avenue, Tucson AZ 85748, USA; kolsen@noao.edu}

\altaffiltext{1}{Visiting astronomer, Cerro Tololo Inter-American Observatory (CTIO), a division of the National Optical Astronomy Observatory, which is operated by the Association of Universities for Research in Astronomy, Inc., under cooperative agreement with the National Science Foundation.}

\begin{abstract}
The yellow supergiant content of nearby galaxies provides a critical test of massive star evolutionary theory. While these stars are the brightest in a galaxy, they are difficult to identify because a large number of foreground Milky Way stars have similar colors and magnitudes. We previously conducted a census of yellow supergiants within M31 and found that the evolutionary tracks predict a yellow supergiant duration an order of magnitude longer than we observed. Here we turn our attention to the SMC, where the metallicity is 10$\times$ lower than that of M31, which is important as metallicity strongly affects massive star evolution. The SMC's large radial velocity ($\sim$160 km s$^{-1}$) allows us to separate members from foreground stars. Observations of $\sim$500 candidates yielded 176 near-certain SMC supergiants, 16 possible SMC supergiants, along with 306 foreground stars and provide good relative numbers of yellow supergiants down to 12$M_\odot$. Of the 176 near-certain SMC supergiants, the kinematics predicted by the Besan\c{c}on model of the Milky Way suggest a foreground contamination of $\leq$4\%. After placing the SMC supergiants on the H-R diagram and comparing our results to the Geneva evolutionary tracks, we find results similar to those of the M31 study: while the locations of the stars on the H-R diagram match the locations of evolutionary tracks well, the models over-predict the yellow supergiant lifetime by a factor of ten. Uncertainties about the mass-loss rates on the main-sequence thus cannot be the primary problem with the models.

\end{abstract}


\section{Introduction}
\label{INTRO}
Yellow supergiants (F0 - G9 I) are evolved, helium burning massive stars that represent a short time period in massive stars' lives as they pass from the blue side of the Hertzsprung-Russell diagram (HRD) to the red supergiant stage or from the red back to the blue. As such, these stars are extremely rare: while the Andromeda Galaxy (M31) contains $\sim$25,000 unevolved (OB-type) stars more massive than 20$M_\odot$ (Massey 2009), Drout et al.\ (2009) estimated only $\sim$16 yellow supergiants in this mass range. 

Identifying a complete sample of yellow supergiants in a galaxy proves extremely useful as they provide a good test of the stellar evolutionary theory. As Kippenhahn \& Weigert (1990) put it, ``[The yellow supergiant] phase is a sort of magnifying glass, revealing relentlessly the faults of calculations of earlier phases." Besides increasing our understanding of massive stars, having reliable evolutionary tracks is vital for interpreting the spectra of distant galaxies using population synthesis codes such as STARBURST99 (Leitherer et al.\ 1999, Vazquez \& Leitherer 2005). While STARBURST99 is a powerful tool, it is no better and no worse than the evolutionary models on which it is based. Additionally, these models are necessary for determining the initial mass function in mixed-age populations, where the tracks are needed both to convert luminosities to masses, and to correct for the ages of stars at each point in the HRD. 

However, the evolutionary models of massive stars require observational testing, as the details of how convection, rotational mixing, and mass-loss are treated all greatly influence the evolutionary calculations(see, for example, discussion by Przybilla et al.\ 2010; Maeder \& Meynet 2008; Maeder et al.\ 2008). In particular, the mass-loss rates on the main-sequence are probably a factor of 3 or more lower than what has usually been assumed in evolutionary models, due to the recent discovery of the importance of including non-homogeneity in the stellar winds (``clumping"), as described by Fullerton et al.\ (2006) and Puls et al.\ (2009), among others. In addition, the red supergiant mass-loss rates depend upon the assumption of uncertain gas-to-dust ratios (Josselin et al.\ 2000, van Loon 2007). 

Drout et al.\ (2009) examined the yellow supergiant content of M31 and found that its relative number as a function of luminosity differed dramatically from the predictions of the evolutionary models. Drout calculated a lifetime based on the number of yellow supergiants compared to unevolved stars, and found that the models predicted lifetimes over an order of magnitude higher. In this study, we extend this test to the SMC where the metallicity is approximately $10\times$ lower than in M31 ($\log$ O/H + 12 = 8.1 versus 9.1, according to Russel \& Dopita 1990 and Zaritsky et al.\ 1994, respectively). This difference in metallicity drastically affects the models via mass-loss rates, which are driven by radiation pressure in highly-ionized metal lines. By comparing the results between M31 and the SMC, we can thus determine if the primary problem with the M31 yellow supergiant lifetimes is how main-sequence mass-loss is treated. 
 
\begin{figure}
\epsscale{0.55}
\plotone{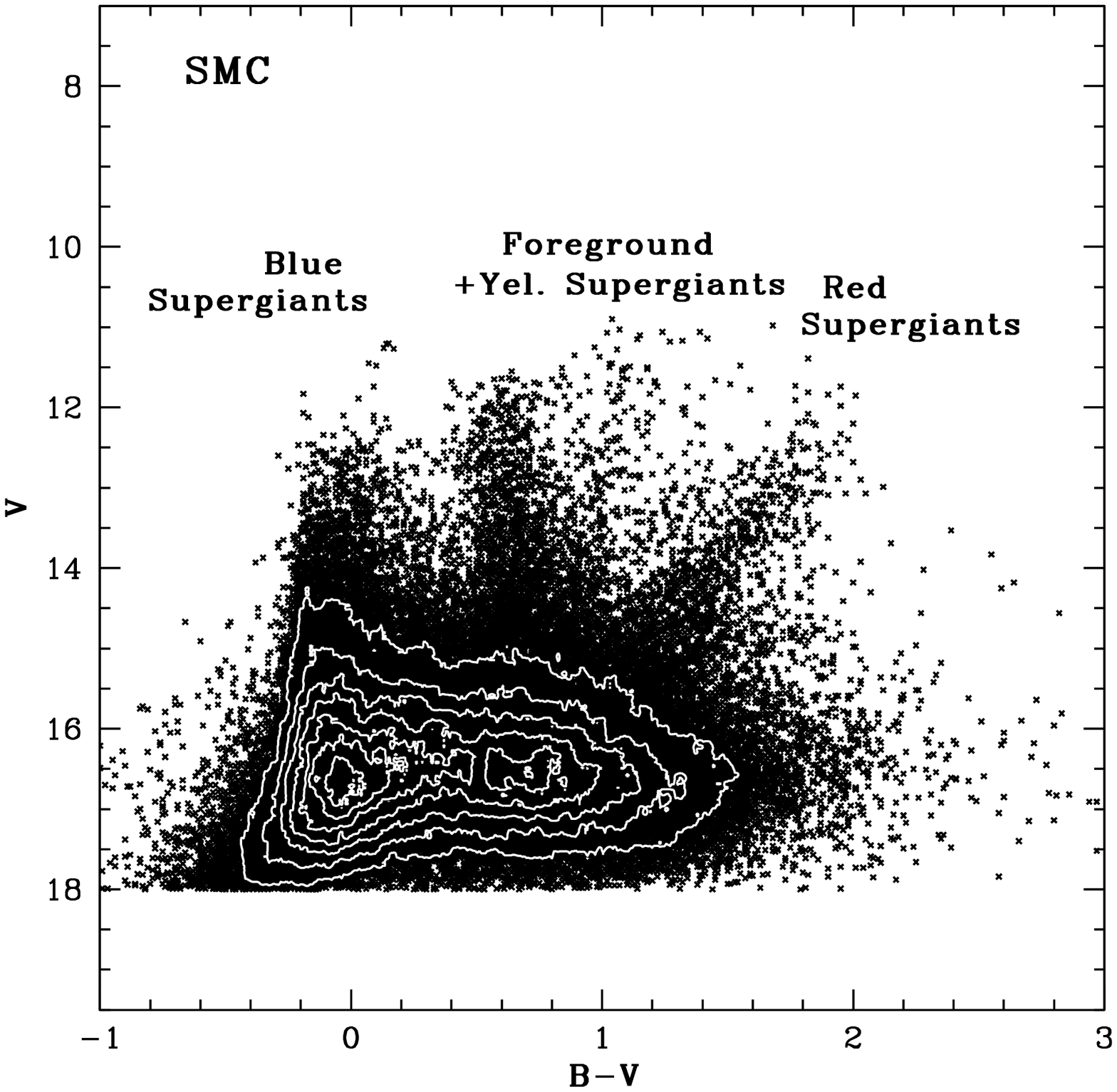}
\epsscale{0.55}
\plotone{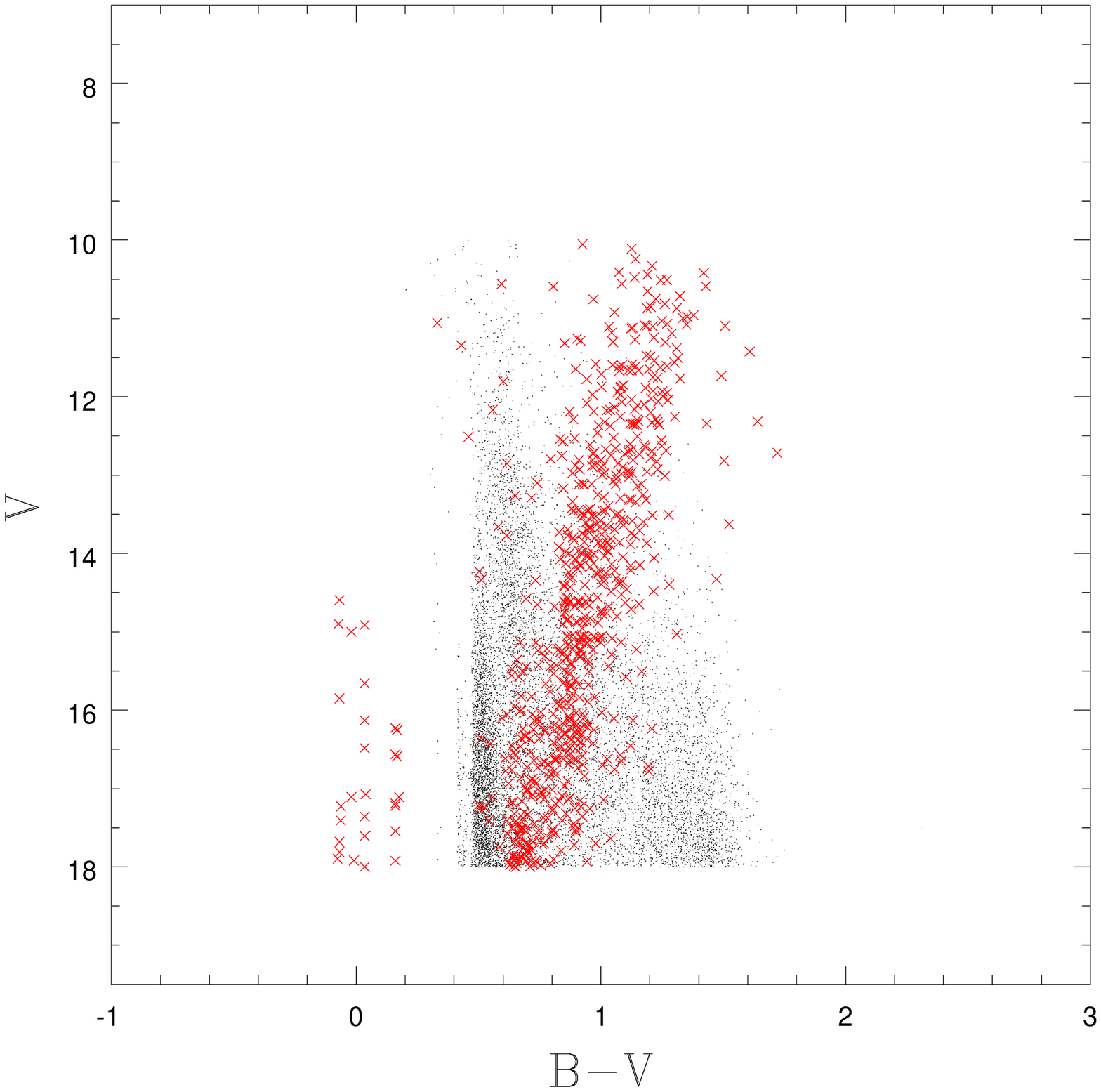}
\caption{\label{fig:smcCMD} Foreground contamination of yellow supergiants. \emph{(upper)} The color magnitude diagram of the SMC is shown, where the data is taken from Massey (2002). \emph{(lower)} The color magnitude diagram of the foreground contamination is calculated using the Besan\c{c}on model (Robin et al.\ 2003) using the same 7.2$^\circ$ area centered at the same galactic latitude and longitude as the SMC. The black dots are dwarfs while the red $\times$s are giants. A comparison between the \emph{upper} and \emph{lower} diagrams show the region where foreground contamination abounds.}
\end{figure}

Although yellow supergiants are among the brightest stars in a galaxy, identification is difficult because they are masked by the presence of foreground stars. This is displayed in Figure~\ref{fig:smcCMD} where Figure~\ref{fig:smcCMD} \emph{upper} shows the SMC's HRD and Figure~\ref{fig:smcCMD} \emph{lower} shows the predicted location of foreground stars from the Besan\c{c}on models of the Milky Way (Robin et al.\ 2003). While the red and blue supergiant regimes are relatively uncontaminated, the yellow supergiants occupy the same colors as the foreground stars. (The comparison also suggests that the model predicts too many foreground stars, as the density of stars is higher in the yellow region in Figure~\ref{fig:smcCMD} \emph{lower} than in Figure~\ref{fig:smcCMD} \emph{upper}.) Because of foreground contamination, we must use some other method to separate the yellow supergiants from the foreground yellow stars. We expect to be able to do this based on radial velocities, as the SMC has a heliocentric radial velocity of 158 km s$^{-1}$ (Richter et al.\ 1987), while stars in the Milky Way should have radial velocities around 0 km s$^{-1}$.

Once the yellow supergiant content has been determined, the SMC's massive star population will have been characterized from one side of the HRD to the other. Previous studies tell us much about OB stars and their evolved descendants, the red supergiants and Wolf-Rayet stars, and thus their numbers are all relatively well known (Massey 2002, Massey et al.\ 1995, Massey et al.\ 2003, Massey \& Duffy 2001, Massey \& Olsen 2003, Mokiem et al.\ 2006). The completion of this massive star survey will provide a testing ground for both current and future stellar models.

In the following sections we will explain how we established membership of yellow supergiants in the SMC and how well our observations matched the evolutionary tracks. In Section~\ref{OS} we describe our observation and reduction procedures. In Section~\ref{A} we discuss how we separated the foreground stars from the SMC supergiants. In Section~\ref{HRD} we put the SMC members on the HRD and compare the results with current evolutionary tracks and in Section~\ref{C} we summarize our findings and list future goals.

\section{Observations and Reductions}
\label{OS}

\subsection{Target Selections}
\label{TS}
To identify F and G supergiants in the SMC, we initially selected stars from the USNO CCD Astrograph Catalogue Part 3 (UCAC3) to have negligible proper motions (less than 15 mas~year$^{-1}$ in $\alpha$ and $\delta$). We chose a 1.75$^\circ$ radius circle around $0^{h}55^{m}11^{s}$ $-72^\circ57\arcmin00\arcsec$ (J2000) to include most of the SMC's optical body and the cataloged OB associations (Hodge 1985). We additionally included a small region centered on NGC 602 ($1^{h}26^{m}40^{s}$ $-73^\circ21\arcmin00\arcsec$ (J2000)) in the wing of the SMC, a region rich in OB stars. For control fields, we selected two 1.75$^\circ$ radius regions at the SMC's galactic latitude ($-44.2^\circ$) but 7.5 degrees higher and lower in galactic longitude ($23^{h}47^{m}41^{s}$ $-72^\circ04\arcmin50\arcsec$ (J2000) and $2^{h}07^{m}43^{s}$ $-71^\circ41\arcmin20\arcsec$ (J2000)). 

\begin{figure}
\epsscale{1}
\plotone{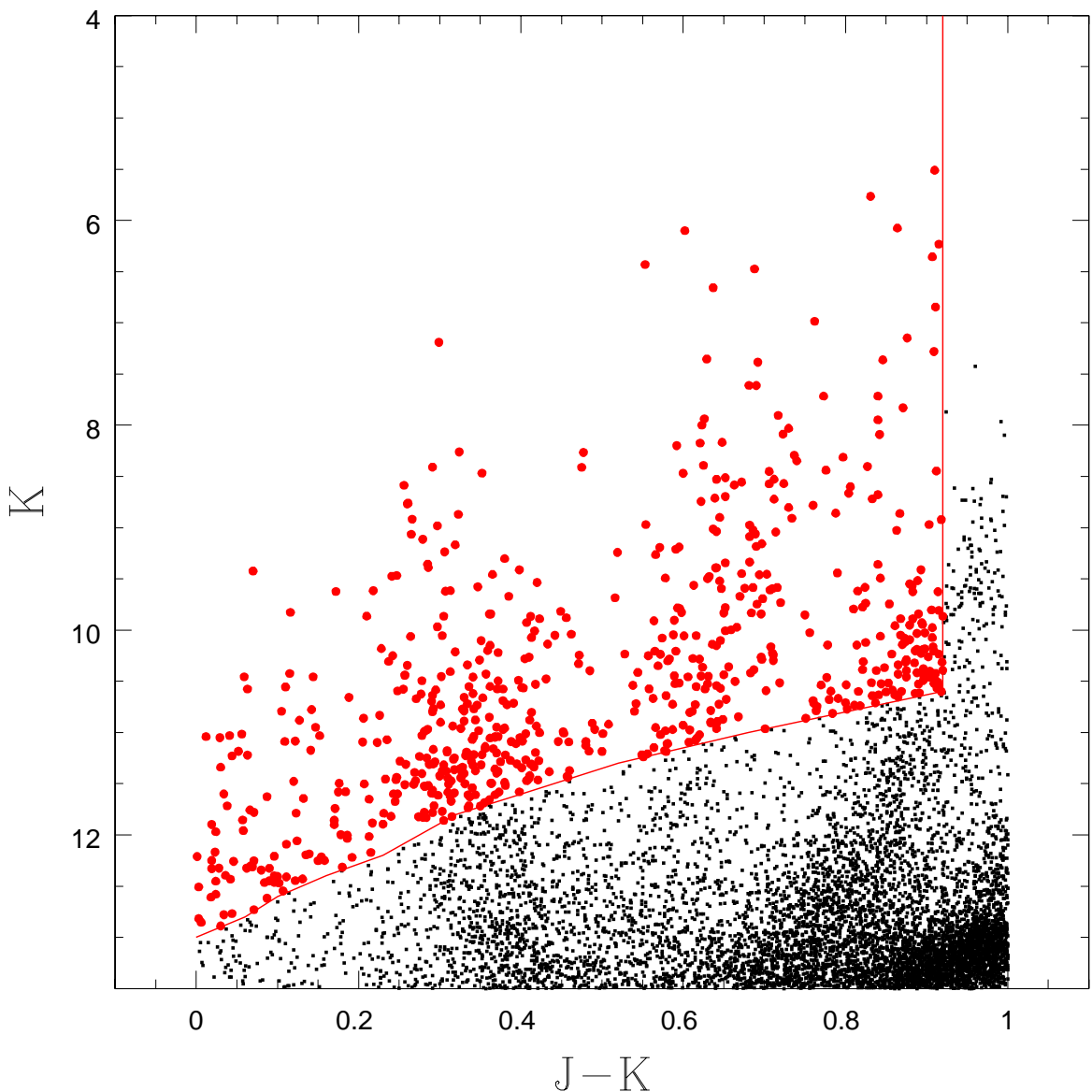}
\caption{\label{fig:2colorsmc} Color selection criteria of program targets. This figure shows the magnitude and color cut-off criteria we used when selecting our program targets. The vertical line is at $J-K$ = 0.92 and the sloped line shows that when $J-K = 0$, stars with $K$ magnitudes fainter than 13.0 were rejected and when $J-K = 0.92$, stars with $K$ magnitudes fainter than 10.6 were rejected. All stars above the sloped line and to the left of the vertical line became our selected program targets. The selected stars appear as red circles and the unselected stars appear as black dots.}
\end{figure}

We used the stars' 2MASS photometry (Skrutskie et al.\ 2006) to then select a sample in the necessary color and magnitude range in order to be complete for yellow supergiants down to 12$M_\odot$. (Although $B-V$ would be more sensitive to $T_{\rm eff}$, reliable $B-V$ values were not readily available for all of the UCAC3 stars.) Following Drout et al.\ (2009), we define the yellow supergiant $T_{\rm eff}$ range as 4800 K to  7500 K. The Geneva evolutionary tracks (Maeder \& Meynet 2001) and the $J$ and $K$ magnitudes of Kurucz's (1992) ATLAS9 stellar atmosphere models then allowed us to define our limiting $K$ magnitude as a function of $J-K$, which is shown in Figure~\ref{fig:2colorsmc}. Additionally, we excluded stars that had low 2MASS color quality codes, along with possible galaxies, clusters and double stars as per the UCAC3. We ended up with 661 possible SMC supergiants and an additional 16 stars in the NGC 602 field for a total of 677 stars. 

\subsection{Observations}
\label{Obs}
\begin{figure}
\epsscale{1}
\plotone{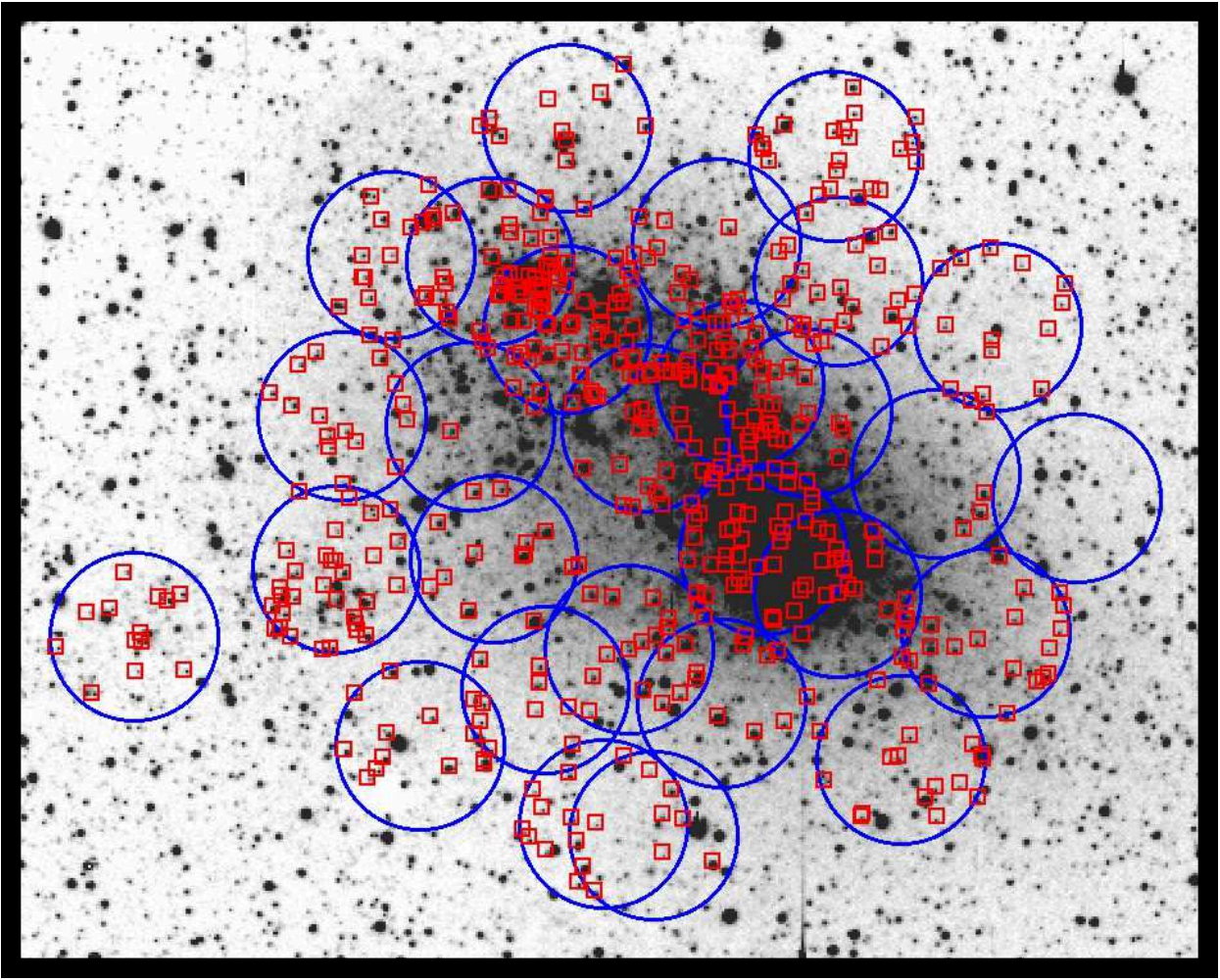}
\caption{\label{fig:SMCobserved} Locations of selected targets. The circles represent our 26 selected SMC field configurations along with the NGC 602 field located in the bottom left corner. Each circle is thus hydra's FOV, 2/3$^\circ$. The boxes highlight the 498 stars we observed and collected usable spectra for. The background image was obtained using the ``parking lot" camera by Greg Bothum.}
\end{figure}

Our observations were taken on the Cerro Tololo 4-meter telescope using Hydra, a multi-object spectrometer with 138 fibers and a 2/3$^\circ$ field of view. Before observing, we created assignment files that matched our targets with specific 2$\arcsec$ diameter fibers on the instrument. More targets were assigned to higher priority fields so if bad weather struck, we would observe the highest priority fields first. In the end, we were able to assign 89.6\% (592) of the stars in 26 SMC fields and 87.5\% (14) of the stars in the NGC 602 field (after being limited by the number of fields we were likely to be able to observe). Additionally, 105 stars were assigned twice in different fiber configurations. The locations of the fields and stars for which we eventually collected spectra are shown in Figure~\ref{fig:SMCobserved}.

As discussed in Section~\ref{INTRO}, we planned on distinguishing SMC supergiants from foreground stars by measuring the stars' radial velocities. The Ca II triplet ($\lambda\lambda\ 8498, 8543, 8662$) is ideal for such a measurement because its strong lines are measurable over a broad temperature range. Our highest priority goal was to obtain red spectra of our objects. But we also aimed to collect blue spectra for later classification purposes.

All of our observations were taken over a cloudy five night span in 2009 October. Two nights were completely overcast and during the other three nights we experienced heavy cirrus. The seeing averaged around 1$\arcsec$, which was acceptable considering the Hydra fibers are 2$\arcsec$ in diameter. In the blue, our wavelength range was 3650 -- 4525 \AA\ with a spectral resolution of 1.3 \AA\ (3 binned pixels). In the red, our wavelength range was 7300 -- 9050 \AA\ with a spectral resolution of 2.6 \AA\ (again 3 binned pixels). For all of our observations we used the same grating, the KPGL-D, and simply changed the blocking filters between the blue (BG39) and the red (OG515). While we aimed at observing all 26 fields in both the red and the blue, after the first eight fields, we quickly realized that due to the first night being lost to weather and the never-ending cirrus, our ambitions were too high. Additionally, we were observing during the Full Moon which impacted the quality of our blue spectra. So, we settled with observing only one control field and the remaining 18 fields in only the red. All of the fields were observed for three consecutive exposures of five minutes each. Additionally we observed four Geneva radial-velocity standard stars (HD154417, HD197076, HD42807, and HD6655) in the red for use as cross-correlation templates. These exposures were 30 seconds long (except for HD154417 which was 15 seconds long) and were taken throughout the run.

\subsection{Reductions}
\label{Red}
Our data were first overscan-subtracted and trimmed and then the remaining bias structure was removed using a master averaged bias. The spectra were then extracted using an optimal extraction algorithm within the IRAF ``hydra" package. During our observing run, we took short He-Ne-Ar comparison arcs and quartz projector flats adjacent to both the blue and red exposures. The comparison arc helped us determine the wavelength scale for the individual fibers while the projector flat helped us determine the location of the spectra on the CCD while also providing the flat fields. Our wavelength calibrations had uncertainties of 0.03 \AA\ in the red and 0.01 \AA\ in the blue. Sky subtraction proved to be difficult because the projector flat lamps didn't illuminate the field uniformly. Therefore, we were unable to remove the fiber-to-fiber transmission variations and the vignetting function very well. However, in the red, the sky contribution was minimal but a few blue spectra were under-exposed and thus discarded. After the data were sky-subtracted, extracted, and wavelength calibrated, we combined the three spectra for each star and rejected the deviant pixels using the IRAF ``avsigclip" algorithm. 

In the end we were able to collect usable spectra for 498, or 74\%, of our selected program targets.

\section{Assigning Membership}
\label{A}
Our ultimate goal was to put stars on the HRD and compare the results to the current evolutionary tracks. But, we first had to determine which stars are SMC members. We did this primarily by measuring radial velocities, and for those stars whose measurements gave inconclusive results, we looked at OI $\lambda$7774 line strengths.

\subsection{Radial Velocity Measurements}
\label{VradMeasure}
To determine the stars' radial velocities, we cross-correlated the spectra using the Ca II triplet. Before cross-correlation, we first normalized the spectra using an 11th order cubic spline and subtracted 1.0 to remove the continuum. During our observing run, we observed four F and G radial velocity standards and after cross-correlating these stars against each other and computing their heliocentric corrections, the results were consistent to $\sim$1 km~s$^{-1}$. We then used the IRAF package ``fxcor" with a wavelength range of 8400 -- 8700 \AA\ (to include the Ca II triplet) to compute the cross-correlations and determine the mean radial velocity and Tonry and Davis (1979) $r$ parameter for each observed program target. These results, along with other identifying information about the program targets, is shown in Table~\ref{tab:allobs}. 

The $r$ parameter measures the degree of the cross-correlation, for which a larger value indicates a more reliable result.  We next investigated the spectra of stars with small $r$ parameters ($r < 25$). Almost all of them were stars too early to show the Ca II triplet and thus cross-correlation using our original wavelength range yielded unsatisfactory results. Because these stars are of early type (OB), their colors were closer to the blue uncontaminated region of the color-magnitude diagram. While we were almost certain of their supergiant status, we still wanted measurements of their radial velocities for confirmation. To do this, we used the Paschen hydrogen lines from P11 to P19. After measuring the velocities of the Paschen hydrogen lines for a few of the spectra, we then used these spectra as templates to cross-correlate the remaining spectra. Once again we used ``fxcor" but this time with a wavelength range of 8400 -- 8900 \AA\ (to include the Paschen lines). 

We confirmed that our program targets were well matched by cross-correlation and our average $r$ parameter was an impressive 75.7. For comparison's sake, when completing this study in M31 (Drout et al.\ 2009), the average $r$ parameter was 33 whereas Tonry and Davis (1979) present single digit values in their study of galaxies. 

While the $r$ parameter is one method of determining error, we also examined the internal and external precision of our computed radial velocities. A comparison between the $r$ parameters and the errors in the radial velocity fits shows that for stars with large $r$ parameters ($r > 130$), velocity errors were around 1.5 km s$^{-1}$. For medium $r$ parameters, errors were around 2~km~s$^{-1}$. And for smaller $r$ parameters ($r < 60$), errors were between 3.5 and 5~km~s$^{-1}$. We were also able to estimate the external precision by comparing the radial velocity results of the 63 stars observed twice. The difference between the two radial velocity results was around 2~km~s$^{-1}$ with a small uncertainty. Given our large $r$ parameter values, we concluded that our velocity accuracy wasn't limited by the exposure level of our spectra, but rather by the stability of the bench mounted spectrograph and our ability to fit the comparison lines, which had residuals near 1~km~s$^{-1}$ as stated above.

Now that we were confident with our radial velocity results, we made a first pass at identifying our program targets as either SMC members or foreground stars. As Figure~\ref{fig:radVelRplot} shows, a first cut at classifying the stars was relatively simple. Just as we had hoped, the plot of radial velocity vs.\ the $r$ parameter is clearly bimodal with one grouping of stars centered around the SMC's radial velocity (158 km s$^{-1}$) and another grouping of stars centered around 0 km s$^{-1}$, the radial velocity of the Milky Way in the direction of the SMC. In general, the offsets are large enough to accommodate velocity dispersion in each galaxy and still remain clearly separated. But, what about the small number of stars ($\sim$30) in the middle? 

\subsection{OI $\lambda$7774 Strengths}
\label{oi}
For stars with intermediate radial velocities in Figure~\ref{fig:radVelRplot}, we used a luminosity sensitive line, OI $\lambda$7774, to determine membership. According to Osmer (1972), OI $\lambda$7774 is strong in F supergiants as compared to its strength in dwarfs due to non-LTE effects and as a result of spheroidicity (Przybilla et al.\ 2000), at least at Galactic metallicities. So, the ``questionable" stars with measurable amounts of OI $\lambda$7774 should be supergiants, while the stars that contained very little OI $\lambda$7774 should be foreground stars. However, this luminosity dependence hasn't yet been tested at SMC metallicities or for the cooler (G-type) supergiants. So, before applying this rule, we first needed to check these two points using the stars clearly separated as a test of the method. We did this by determining the effective temperature and luminosity for every observed star using the relations described below in Section~\ref{makeHRD}, and measure the equivalent width of the OI $\lambda$7774 line. We assume that stars with very large radial velocities are SMC supergiants and those with low velocities are foreground dwarfs. 

The results of measuring the equivalent widths of the OI $\lambda$7774 lines are shown in Table~\ref{tab:derived}. While we knew that OI $\lambda$7774 is strong in F supergiants (Osmer 1972), we didn't know its behavior for G-type supergiants. We first needed to determine if there is a temperature cut-off for the luminosity dependence. Indeed, after examining our results, we determined that the relationship is only significant in hotter stars, specifically our category 1 stars with $\log T_{\rm eff} > 3.72$ ($ > $5200 K) at SMC metallicities ($z = 0.2z_\odot$). So, almost all of our category 1 supergiants with $\log T_{\rm eff} > 3.72$ had a measurable amount ($ > 0.2$\AA) of OI $\lambda$7774 while almost all of our stars with $\log T_{\rm eff} \leq 3.72$ didn't. This lower temperature limit of 5200 K falls near the bottom of our identified yellow supergiant temperature range of 4800 to 7500 K (Drout et al.\ 2009). While this luminosity dependence appears to hold true for most G supergiants, it may not hold true for the coolest of them. For the category 1 stars with higher temperatures, we confirmed that we could use the OI $\lambda$7774 line as a method of determining membership and for those that we couldn't use the OI $\lambda$7774 line, we assigned those to category 2 stars. 

\subsection{Membership Determination}
\label{md}
\begin{figure}
\epsscale{1}
\plotone{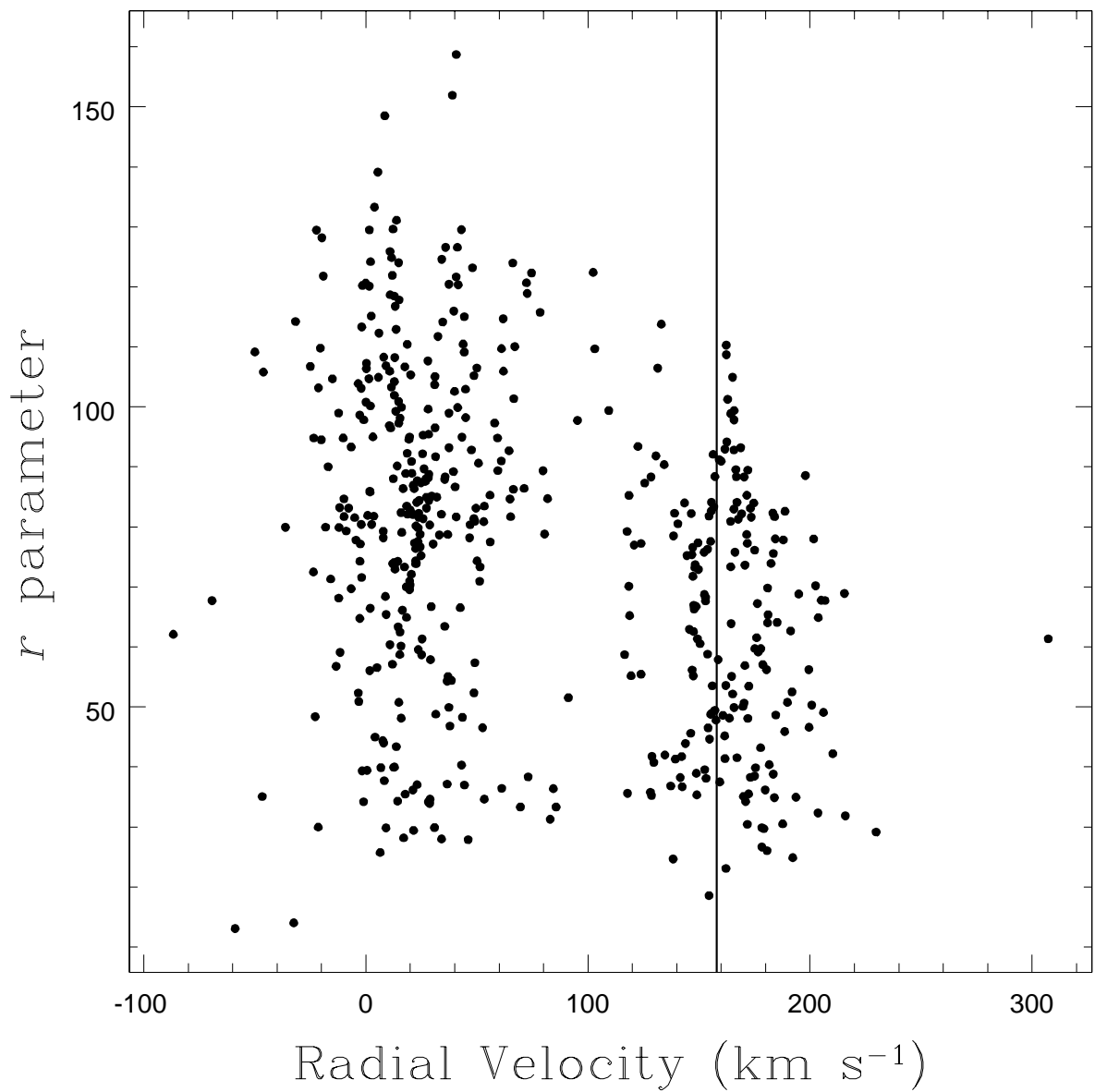}
\caption{\label{fig:radVelRplot} Radial velocity results. This bimodal distribution clearly displays the separation between the foreground stars (centered around 0 km s$^{-1}$) and the SMC supergiants (centered around 158 km s$^{-1}$ shown by the vertical line). Since this is a plot of the $r$ parameter vs.\ radial velocity, as the y values increase, so does the reliability of the results. But even at the lowest $r$ parameter values, our velocity errors were between 3.5 and 5 km s$^{-1}$.}
\end{figure}

\begin{figure}
\epsscale{0.48}
\plotone{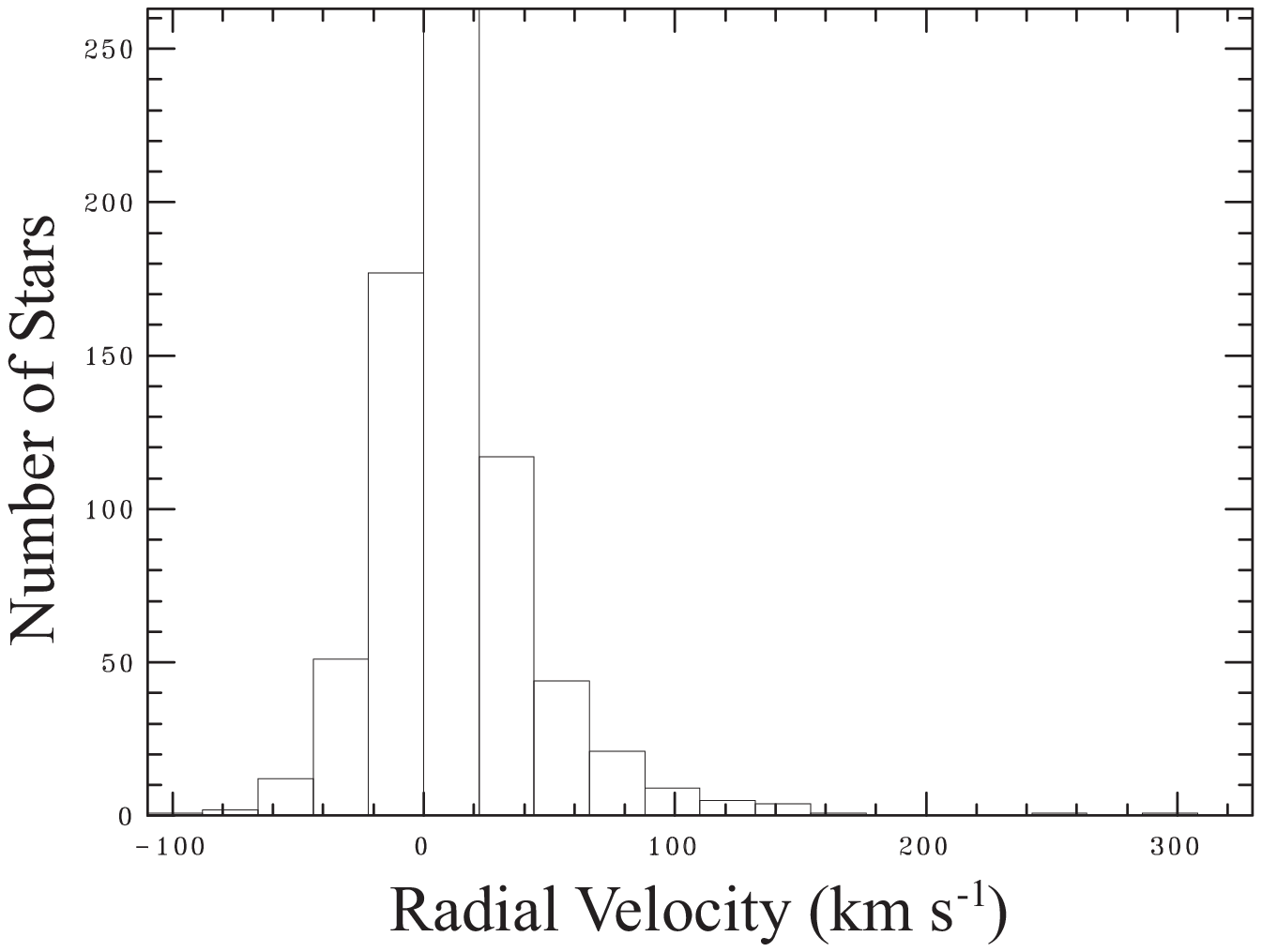}
\epsscale{0.48}
\plotone{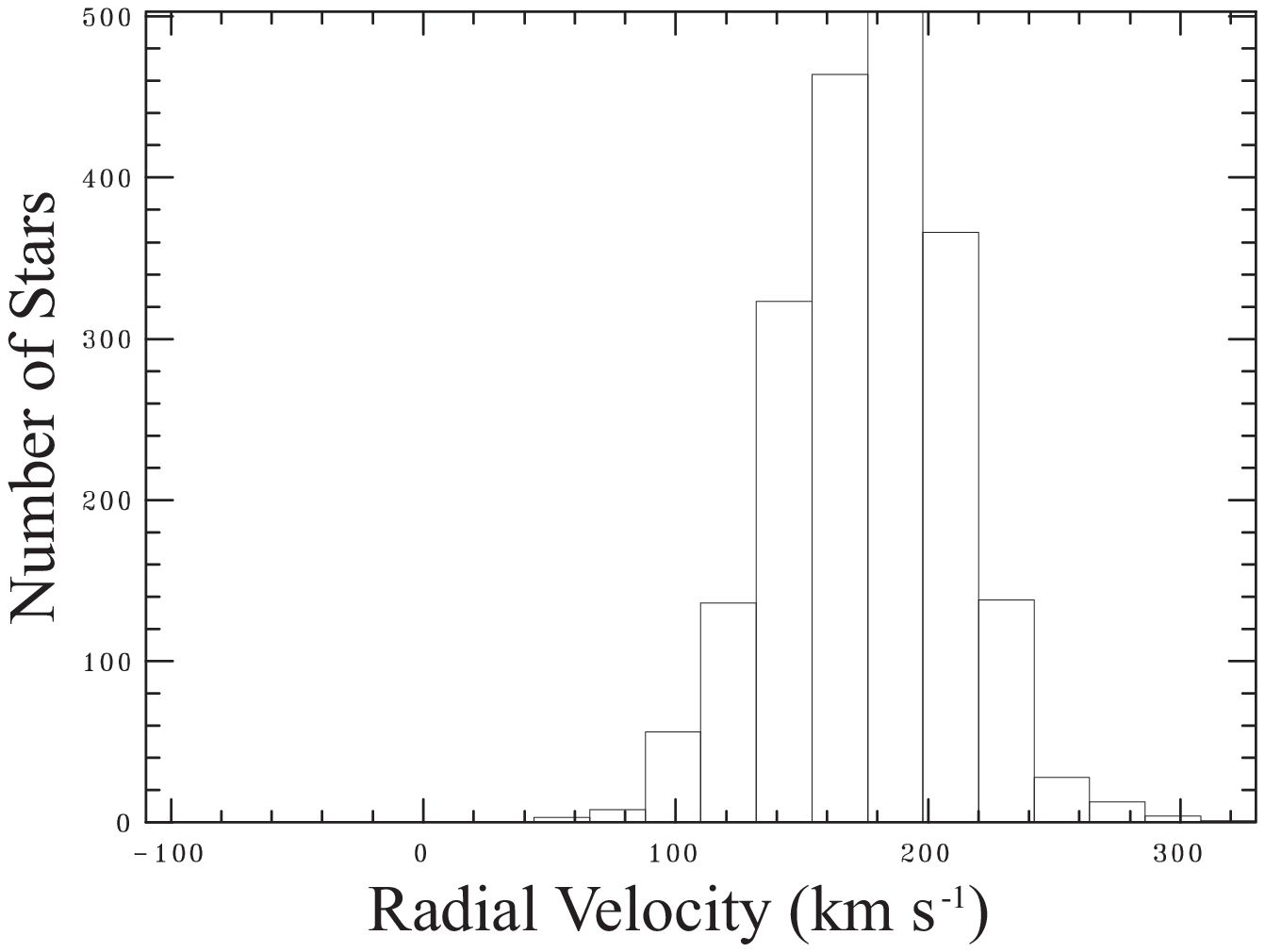}
\epsscale{0.8}
\plotone{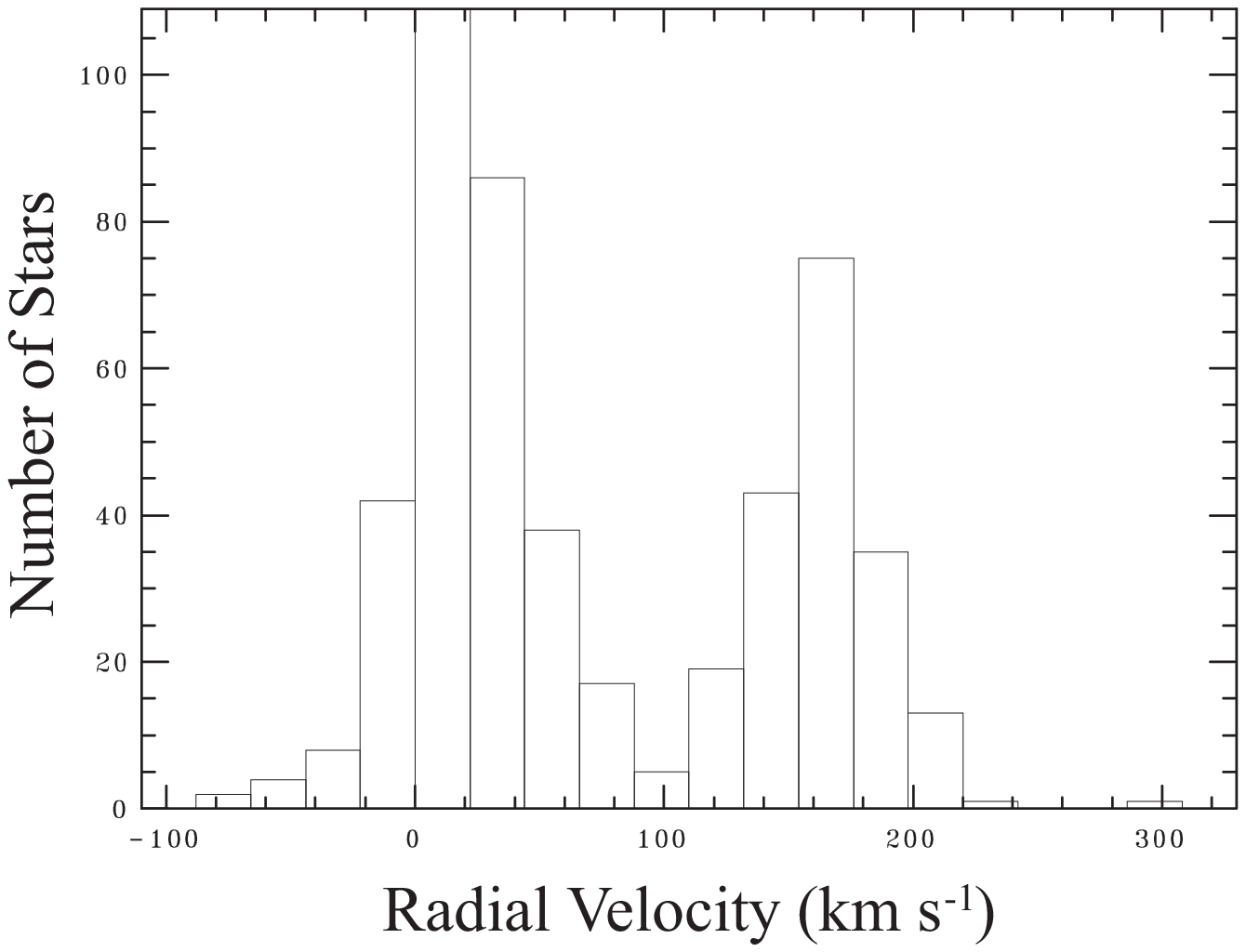}
\caption{\label{fig:histograms} Radial velocity histograms. The \emph{upper left} figure represents the results of the Besan\c{c}on model for foreground stars in the same location as our observed field. The \emph{upper right} figure represents the radial velocity results of Evans \& Howarth (2008) for SMC O, B, A, F and G type stars. The \emph{lower} figure represents our measurements. The y-axes should be considered on a relative basis.}
\end{figure}

Before final membership determination, we need to determine the expected radial velocity range for SMC stars. As seen in Figure~\ref{fig:radVelRplot}, the radial velocity distribution of SMC stars extends to $\sim$240~km~s$^{-1}$ with one star at 300 km s$^{-1}$. Since we are taking the radial velocity of the center of the SMC to be $\sim$160~km~s$^{-1}$, given a Gaussian distribution, a reasonable lower limit would be $\sim$80~km~s$^{-1}$. This value is consistent with the radial velocity histograms for our observed results and the Besan\c{c}on models for the Milky Way shown in Figure~\ref{fig:histograms} (\emph{upper left})where there appears to be a sharp drop in foreground stars above $\sim$90~km~s$^{-1}$. These values, as well as our overall radial velocity results, are also consistent with previous findings by Evans \& Howarth (2008) for SMC members. Figure~\ref{fig:histograms} (\emph{upper right}) shows their radial velocity results for $\sim$2500 O, B, A, F and G type stars in the SMC. This histogram again suggests that the number of SMC stars drastically increases at a radial velocity value of $\sim$80 km s$^{-1}$ and then drops off at a value close to 240 km s$^{-1}$.

\begin{figure}
\epsscale{1}
\plotone{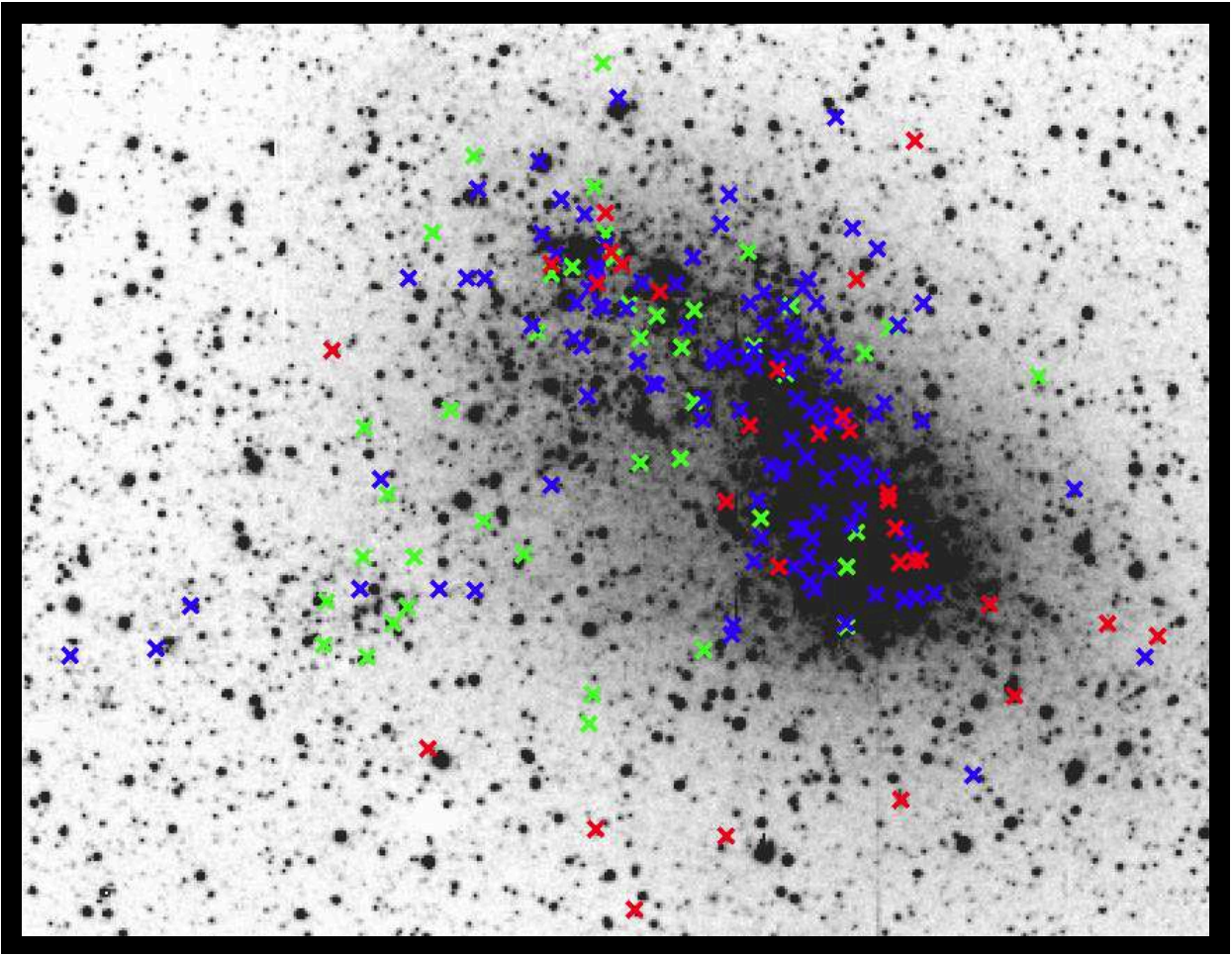}
\caption{\label{fig:VelColor} Radial velocity distribution across the SMC. The red $\times$s represent the stars with low radial velocities (83 -- 138.5 km s$^{-1}$). The blue $\times$s represent medium radial velocities (139 -- 178.4 km s$^{-1}$) and the green $\times$s represent high radial velocities (178.8 and 307.3 km s$^{-1}$). According to Stanimirovi\'{c} et al.\ (2004), the northeast section and the wing have higher velocities than the southwest section. This is consistent with our results because we found more green $\times$s in the upper left hand corner and more red $\times$s in the lower right hand corner of the SMC.}
\end{figure}

The SMC is also rotating and thus stars on one side of the SMC aren't moving at the same velocity as stars on the other side of the SMC. As shown by a study of the SMC's HI kinematics in Stanimirovi\'{c} et al.\ (2004), the northeast section and the wing both have systematically higher velocities than the southwest section. These values typically range between 80~km~s$^{-1}$ and 240~km~s$^{-1}$. Figure~\ref{fig:VelColor} shows how well our radial velocity results fit with these results. Generally we find higher velocity stars in the northwest quadrant and the wing and lower velocity stars in the southwest. However, we will also note that a smattering of low velocity stars are mixed in with the high velocity stars and vice versa which further underscores the inherent complexity of the SMC kinematics. 

Returning to the question of which velocity range to use for membership determination, both Figure~\ref{fig:radVelRplot}, Figure~\ref{fig:histograms}, and Stanimirovi\'{c} et al.\ (2004) suggest that most SMC members have velocities between the extremes of 80 and 240~km~s$^{-1}$.

\begin{figure}
\epsscale{1}
\plotone{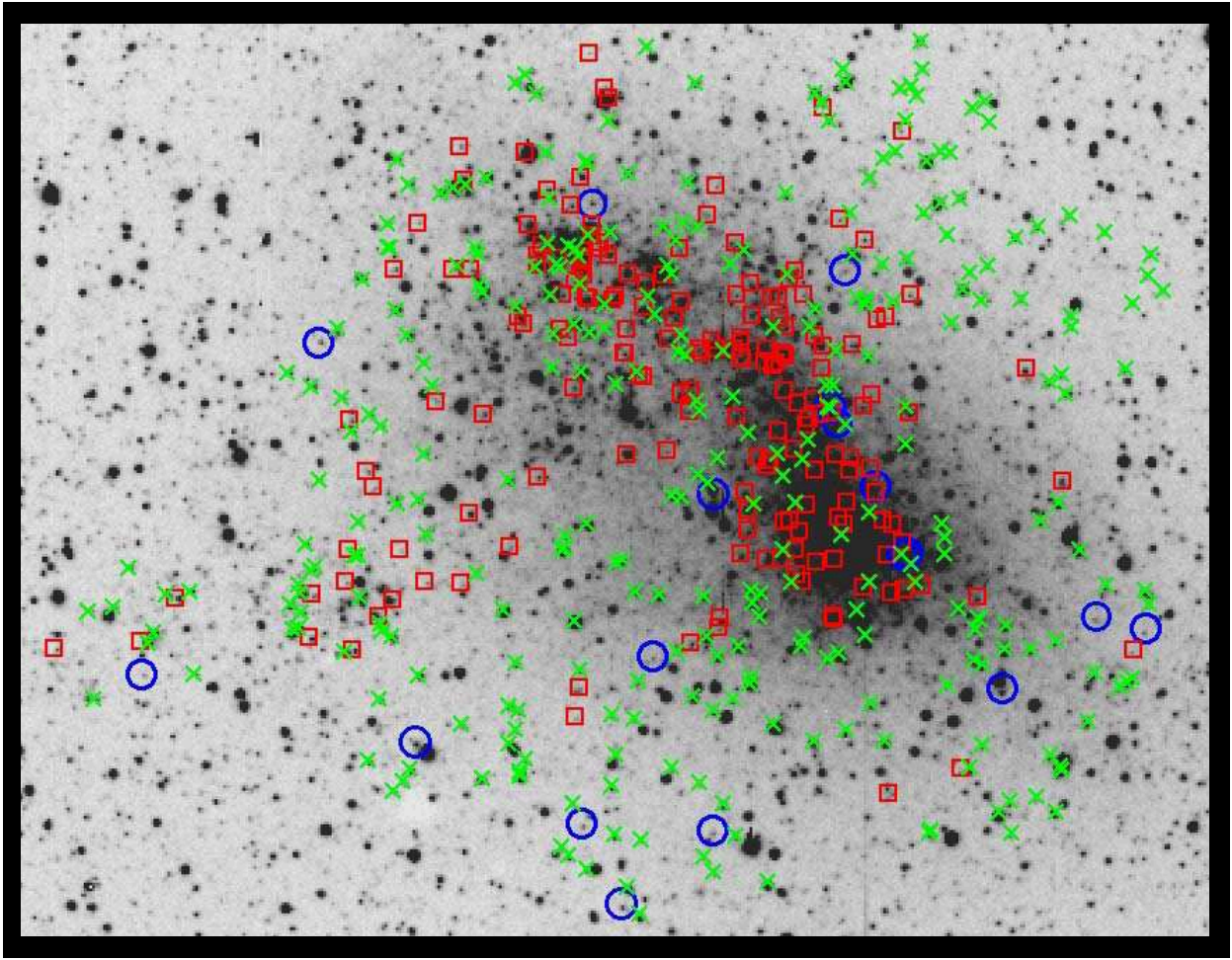}
\caption{\label{fig:SMCcategory123} Locations of observed stars across the SMC. The red boxes represent the 176 category 1 identified supergiants, blue circles represent the 16 category 2 possible supergiants and the green $\times$s represent the category 3 non-members. Notice that some of the supergiants are in the NGC 602 field we identified as having H$\alpha$ emission. Also, while the category 1 boxes are clearly centered around the SMC, the category 2 circles and the category 3 $\times$s are fairly evenly dispersed throughout the selected area. This dispersion is characteristic of foreground stars and thus we are fairly certain that at least a few of our category 2 stars are in fact foreground stars.}
\end{figure}

Based on Figure~\ref{fig:radVelRplot} we can conclude that stars with radial velocities higher than 135 km s$^{-1}$ are SMC supergiants and thus were assigned to category 1. And, as discussed above, stars with radial velocities lower than 80 km s$^{-1}$ are foreground stars and thus were assigned to category 3. We base our classification of the 32 stars in the middle (with radial velocities between 80 km s$^{-1}$ and 135 km s$^{-1}$) on the OI $\lambda$7774 abundances. Of these 32, 16 stars are hotter than $\log T_{\rm eff}$=3.72, and thus are easy to classify based upon the OI $\lambda$7774 line. Fourteen of these sixteen show the OI $\lambda$7774 line, and are clearly SMC supergiants, while two do not, and are classified as foreground stars. The remaining 16 stars are considered likely, but uncertain, supergiants, and are assigned to category 2. We show an updated version of our radial velocity plot in Figure~\ref{fig:radVelRplotMem}, this time with the category 1, 2 and 3 stars now identified. We also show the spatial distribution of the three categories in Figure~\ref{fig:SMCcategory123}. Even though our radial velocity measurements are accurate, the OI $\lambda$7774 line provides useful information when assigning membership for $\sim$20 stars.

\subsection{Contamination}
\label{contam}
We identified 176 category 1 SMC supergiants, 16 category 2 potential SMC supergiants, and 306 category 3 foreground stars, and thus, our overall contamination by foreground yellow stars is between 61\% and 65\%. However, before even sitting down at the telescope, we estimated our expected contamination by comparing the number of stars in each control field with the number of stars in our SMC field (excluding the NGC 602 field). Because these fields are completely populated by foreground stars, and are found at the same Galactic latitude, they provide a direct measure of the expected contamination. This calculation estimated a contamination of 67\%, or essentially the same percentage as our observed contamination. Spectroscopy in one of the control fields revealed no supergiants and the other control field was never observed. 

We additionally studied the contamination using the Besan\c{c}on models to run simulations of how many foreground yellow stars we might expect when looking at the same patch of sky and using the same selection criteria. Partially due to the over-prediction of the number of giants (as discussed in Section~\ref{INTRO}), the Besan\c{c}on models predict nearly 700 foreground stars in the same patch of sky where we found around 300. We are confident that this is an overestimation of the number of disk giants (rather than halo giants), as the Besan\c{c}on models predict only 2 or 3 halo giants in our sample. However, the Besan\c{c}on models also show heavy contamination due to dwarfs. These contamination percentages are well below the 98\% contamination found by Drout et al.\ (2009) in their M31 study. But, this is expected, because the SMC is further away from the galactic equator and the stars in our SMC sample are several magnitudes brighter than those in M31. 

We also looked at the expected contamination by fast moving foreground halo giants in our list of 176 category 1 SMC supergiants. Figure~\ref{fig:histograms} shows radial velocity histograms for our observed results and for the Besan\c{c}on model results. These histograms show that even with a factor of two difference ($\sim$700 vs.\ $\sim$300), we still only estimate eight stars (4\%) in our list of category 1 supergiants to be foreground stars. But, the real answer probably lies closer to three or four stars, or 2\%. As for the category 2 stars, the Besan\c{c}on models predict that nearly all of these stars are actually foreground stars. But, because of the factor of two discrepancy, this model shouldn't be solely relied upon.

\section{H-R Diagram}
\label{HRD}
In this section we will describe the process of transforming the observables into physical properties, the completeness of our survey, and how our results matched the evolutionary tracks.

\subsection{Making the HRD}
\label{makeHRD}
To put stars on the HRD, we first used the star's color to determine its effective temperature. Even though we used $J-K$ colors as the selection criteria, we wanted to use the star's $B-V$ color for the color to temperature transformation since it is much more sensitive to temperature. For the stars without reliable $B-V$ colors we used their $J-K$ colors for the transformations. In general the $V$ magnitudes and $B-V$ colors are from the same source. Some stars do not appear in the Massey (2002) catalogue, and are in the saturated regime of the Zaritsky et al.\ (2002) catalogue, where stars brighter than $V$ = 13.5 are considered unreliable (cf.\ their Section 3.2). These were evident by having nonphysical $B-V$ (and/or $U-B$) colors. In such cases, however, the Zaritsky et al.\ B magnitudes are below their saturation limit ($B < 13.5$). For reasonably isolated stars, we could then adopt the ASAS-3 (All Sky Automated Survey) $V$ magnitude and form the $B-V$ color from the two catalogues. An example is 2MASS J004262977-7158227, where Zaritsky et al.\ shows $V$ = 11.87 with a formal uncertainty of 0.32 magnitudes. Even from the rough photometry in catalogues based on Schmidt plate-scans, such as the \emph{Hubble Space Telescope} Guide Star Catalogue short-$V$ exposure, we can tell this is incorrect. The star shows no comparably bright companions closer than 90$\arcsec$, and thus the ASAS-3 magnitude of $V=13.15\pm0.02$ should be reliable. Using this $V$ and the Zaritsky et al.\ $B$, we get $B-V$ = 0.92, and Zaritsky et al.\ $U-B$ = 0.41, consistent with an ordinary early-K foreground dwarf.

For still brighter stars not appearing in the Massey (2002) catalogue, and lacking other photometry in the literature, we revert to showing only $V$ from ASAS-3. For a few large-amplitude variables, such as the long-period Cepheids, we show the $V$ magnitudes at maximum from either published lightcurves or from inspection of the ASAS-3 timeseries plots. References for the magnitudes and colors of each star can be found in Table~\ref{tab:allobs}.

Now that we had satisfactory colors for each of our stars (whether $B-V$ or $J-K$) we needed to form relationships between color and $T_{\rm eff}$. To do this, we used the ATLAS9 model atmospheres (Kurucz 1992) with reasonable temperatures (4,000 -- 10,000 K), low surface gravities and an appropriate metallicity ($0.3\times$ solar). Before the transformation, we first corrected our colors for interstellar reddening. To correct the $B-V$ colors, we used $E(B-V) = 0.09$ (Massey et al.\ 1995) and for the $J-K$ colors, we used $E(J-K) = 0.15 \times E(B-V)$ (Schlegel et al.\ 1988). We then determined the following equation to calculate the $T_{\rm eff}$ from $(B-V)_0$: $$\log T_{\rm eff}=3.929-0.770(B-V)_0+2.0994(B-V)_0^2-3.7493(B-V)_0^3$$ $$+3.5979(B-V)_0^4-1.7438(B-V)_0^5+0.0333(B-V)_0^6$$ We also determined the following equation to calculate the $T_{\rm eff}$ from $(J-K)_0$: $$\log T_{\rm eff}=3.979-1.490(J-K)_0+6.063(J-K)_0^2-18.277(J-K)_0^3$$ $$+29.931(J-K)_0^4-24.360(J-K)_0^5+7.759(J-K)_0^6$$ For $B-V$, the equation is valid for $-0.01 \leq (B-V)_0 \leq 1.69$ and for $J-K$ the equation is valid for $-0.02 \leq (J-K)_0 \leq 0.92$. The high number of terms was necessary to obtain a smooth fit and a low RMS (0.003 dex). 
	
\begin{figure}
\epsscale{1}
\plotone{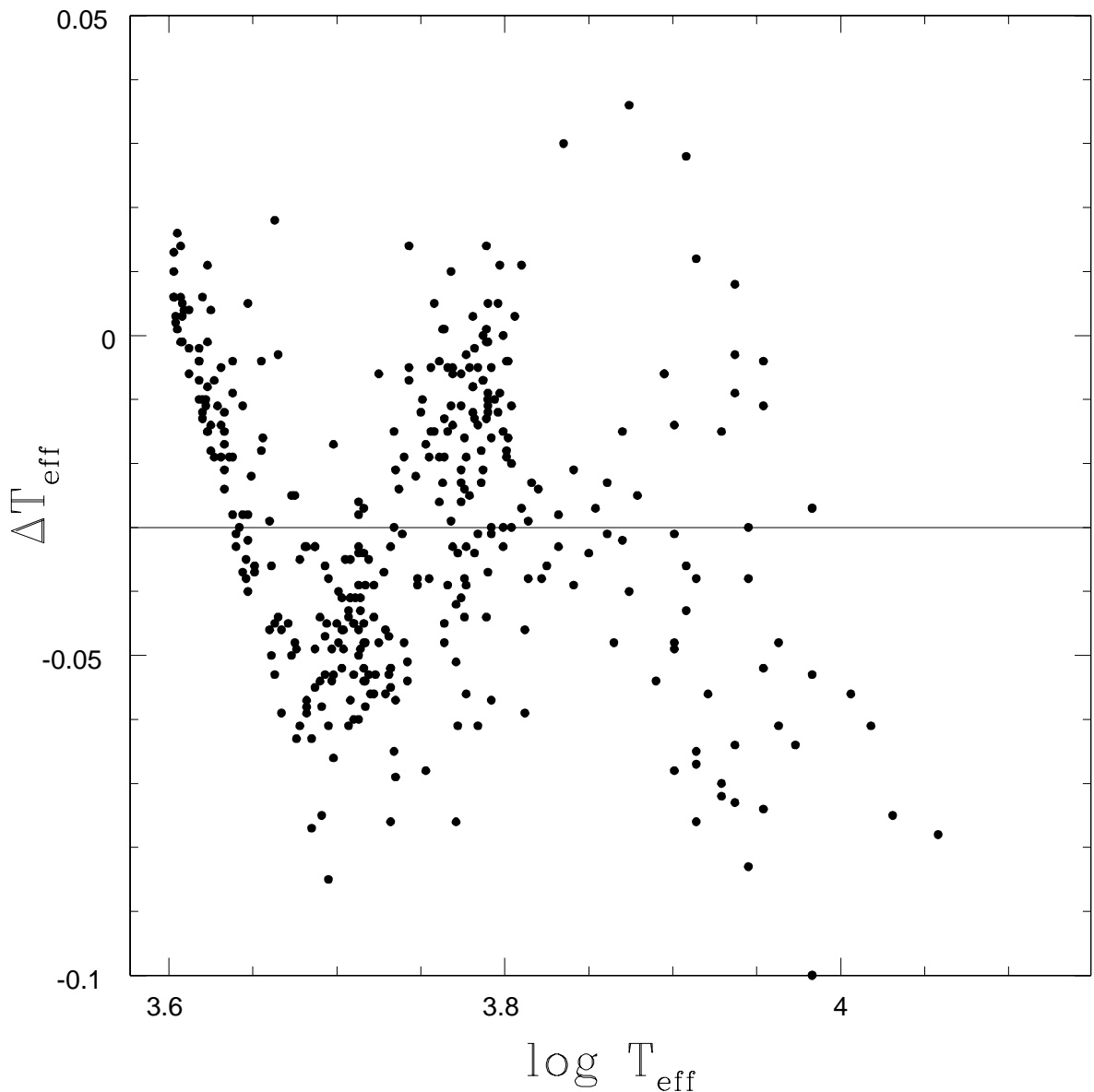}
\caption{\label{fig:tempdiff} Differences between the $T_{\rm eff}$ determined from $B-V$ and $J-K$ for our observed targets. On the x-axis is the $\log T_{\rm eff}$ as determined using the $B-V$ colors and on the y-axis is the difference in $\log T_{\rm eff}$ as determined using $(\log T_{\rm eff (J-K)} - \log T_{\rm eff (B-V)})$. The horizontal line at -0.03 dex is the median systematic difference. For comparison, the expected random error is 0.03 dex.}
\end{figure}

For the above $J-K$ to $\log T_{\rm eff}$ calibration, we used the CIT colors predicted by the ATLAS9 models\footnote{http://kurucz.harvard.edu/grids.html} since their system is well defined. We then transformed the 2MASS $J-K_S$ colors to CIT $J-K$ colors using Carpenter (2001). If one assumes that $B-V$ is uncertain by 0.03 magnitudes and that the reddening may vary by 0.05 magnitudes in $E(B-V)$, then the uncertainty in $\log T_{\rm eff}$ is overshadowed by the uncertainty in reddening and works out to be around 0.03 dex. Similarly, if we assume that the uncertainty in $J-K$ is 0.03 magnitudes, we expect the uncertainty in $\log T_{\rm eff}$ to again be 0.03 dex where the error is primarily dominated by the color uncertainty since the reddening correction is small. As Figure~\ref{fig:tempdiff} shows, there clearly is a \emph{systematic} error in the difference between the $T_{\rm eff}$ computed with the $B-V$ color and with the $J-K$ color. This systematic error is on the order of the expected random error, and we attribute it to small inaccuracies and generalizations present in the ATLAS9 models. For example, the models are LTE and plane parallel (with no stellar winds). In the end we relied on $J-K$ colors to compute the star's effective temperature in only 7 of 192 cases. 

With effective temperatures in hand, we wanted to determine the stars' bolometric luminosities. To do this, we used the ATLAS9 models to form the following relationship between $T_{\rm eff}$ and the star's bolometric correction at $V$: $${\rm BC} = -253.581+131.855\log T_{\rm eff} - 17.1419(\log T_{\rm eff})^2$$ This is valid for $3.603 \leq \log T_{\rm eff} \leq 4.123$. 

The error in the bolometric correction using $V$ is severely dominated by the reddening correction and is around 0.15 magnitudes. We contemplated using $K$ instead because reddening has such little effect, but because $BC_{K}$ is such a steep function of the $T_{\rm eff}$, the errors are virtually identical. We were now able to determine the stars' $\log L/L_\odot$ using the SMC's distance modulus of 18.9 (van den Bergh 2000). The results are shown in Table~\ref{tab:derived}.

\subsection{Completeness}
\label{comp}
Out of our original list of 677 targets, we identified 176 stars as candidate SMC supergiants and an additional 16 stars as probable SMC supergiants. However, many of our targets are bright and thus have been previously classified. As shown in Table~\ref{tab:allobs}, a literature search yielded spectral classifications for 34\% of our program targets. Among these, the stars we identified as SMC supergiants were indeed identified as supergiants in the literature, and the stars we identified as foreground stars were indeed identified as dwarfs in the literature. However, this doesn't hold true on a star-by-star basis. We believe that in most cases the literature luminosity classes are at fault because they are based on spectral lines instead of radial velocity measurements and the incorrect results make use of low-dispersion objective-prism spectra. When determining the classification of an SMC star, the process is made more difficult by the SMC's low metallicity. Thus, stars with weak metal lines may be misclassified as dwarfs when they are in fact SMC supergiants.

In order to further check that our survey did not exclude any of the most luminous members, we examined the Evans \& Howarth (2008) survey of radial velocities and spectral types for a variety of SMC members. However, his list of F- and G- stars begins below our faint magnitude limit (corresponding to 12$M_\odot$) and hence does not contain any highly luminous and massive members. 

Besides looking at literature classifications, we also examined our blue spectra and determined crude spectral types based on prominent metal lines for 209 stars, or 42\% of our observed sample. These results are also shown in Table~\ref{tab:allobs}. Once again, overall we identified our category 1 stars as supergiants and our category 3 stars as dwarfs but for many individual cases this fails. These failures reinforced our distrust in the abilities of spectral types to faithfully discriminate SMC supergiants from foreground dwarfs, particularly for fainter stars. We therefore put no weight on the spectral classifications and didn't use them further.

We were now able to evaluate our study's completeness before placing the stars on the HRD. Firstly, we didn't observe all 677 of our selected targets. But, our literature search only identified six of our unobserved targets as ``known" yellow supergiants. These stars are listed in Table~\ref{tab:knownSG}. Additionally, our literature search turned up 13 ``known" yellow supergiants that didn't make it through our selection criteria. These stars and the reasons they weren't selected are also listed in Table~\ref{tab:knownSG}. But, are these spectral classifications reliable? Five of the stars are fairly bright ($\sim$11th magnitude) and thus we believe their classifications. However, the remaining fourteen stars are closer to 14th magnitude and thus their classifications are much more uncertain. While we acknowledge that our 677 selected targets might not encompass every SMC yellow supergiant, we believe that we're missing only a few, rather than tens of stars.

\subsection{Discussion of the HRD}
\label{DHRD}
\begin{figure}
\epsscale{1}
\plotone{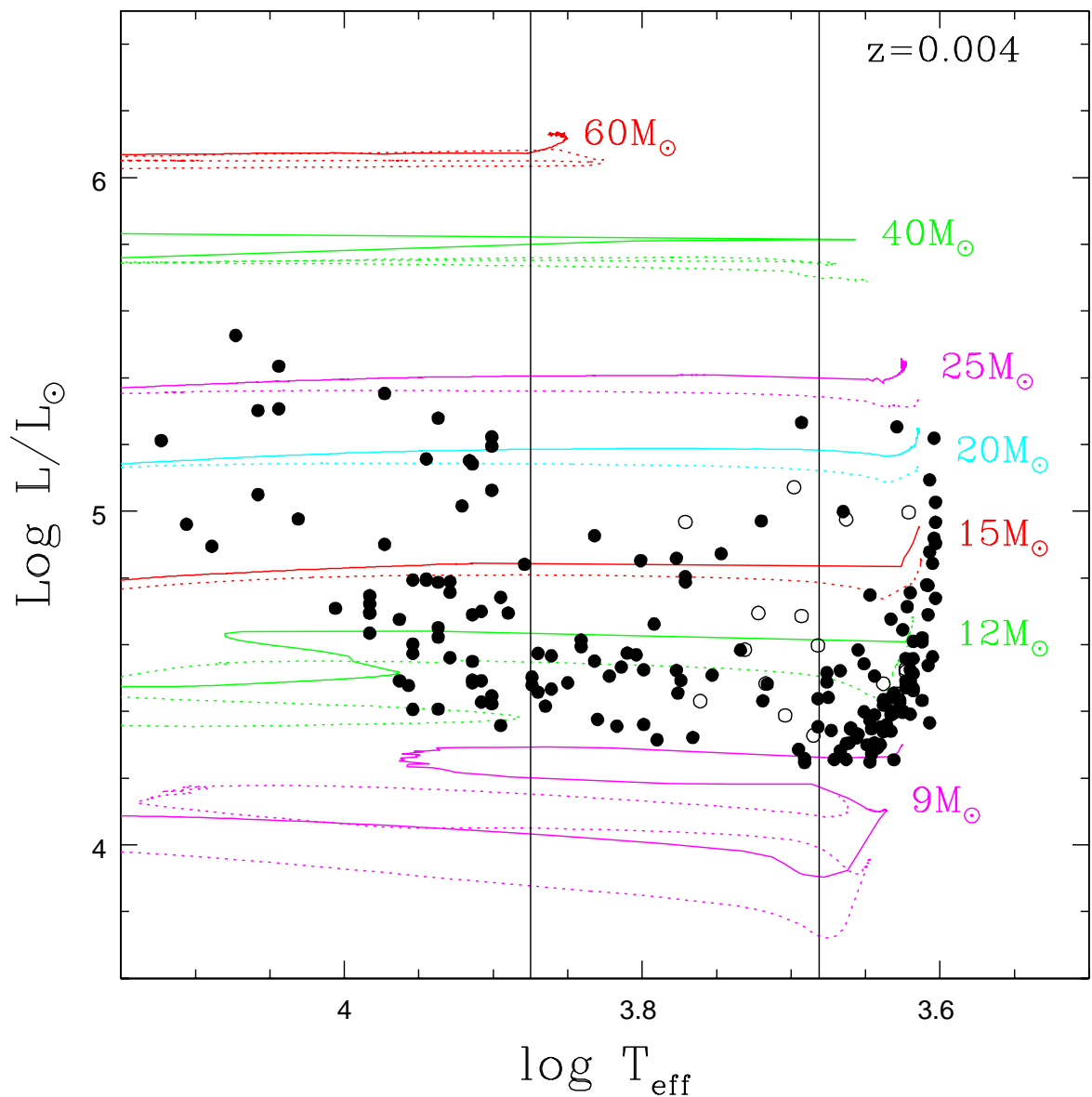}
\caption{\label{fig:SMCfstarsHRD} HRD with yellow supergiant candidates. Solid lines indicate Geneva evolutionary tracks ($z = 0.004$) with an initial rotation of 300 km s$^{-1}$ while the dashed lines indicate no initial rotation. The yellow supergiant region lies between the two solid black vertical lines at 4800 K and 7500 K. Solid circles represent our category 1 SMC supergiants and open circles represent our category 2 possible but not probable SMC supergiants.}
\end{figure}

After determining the effective temperatures and luminosities for our 192 category 1 and 2 stars, we placed our supergiants on the HRD. Our results are shown in Figure~\ref{fig:SMCfstarsHRD} along with the $z = 0.004$ Geneva evolutionary tracks for models with initial rotational velocities of 300 km s$^{-1}$ (solid curves) and 0 km s$^{-1}$ (dashed curves). While we believe that the models with an initial rotation of 300 km s$^{-1}$ are more realistic\footnote{Note that although a rotational velocity of 300 km s$^{-1}$ is higher than usually attributed to early-type stars, this value refers only to the initial equatorial velocity on the zero-age main-sequence. At Galactic metallicities, this would correspond to an average velocity during the main-sequence phase of 180 -- 240 km s$^{-1}$ (Meynet \& Maeder 2003), a value in accord with Galactic O-type stars (Conti \& Ebbets 1977, Penny 1996). The appropriate value to use at SMC metallicities has not yet been fully established; see, for example Penny \& Gies (2009).}, the models with no initial rotation provide an interesting comparison. 

Many of our category 1 and 2 stars fall within the two black lines placed at 4800 K and 7500 K denoting the yellow supergiant region. However, there also appears to be a large group of both red and blue supergiants in our sample. This is due to our lenient color selection criteria, which assumed we included all of the yellow supergiants. An inspection of the supergiants and their corresponding evolutionary tracks confirms that we did a good job at identifying the supergiants down to and well below 12$M_\odot$. 

We can now proceed to the main point of this paper, namely comparing the distribution of yellow supergiants in the HRD with that predicted by the evolutionary tracks. We begin by noting that the overall location of the stars is well matched by the tracks, in that we don't see yellow supergiants at higher luminosities than what the tracks predict. For example, there are not yellow supergiants at $\log L/L_\odot = 6$. But, are their relative numbers correct?

As argued in Section~\ref{comp}, our sample is complete down to 12$M_\odot$. In Table~\ref{tab:numbers} we list the number of yellow supergiants we observed between various mass tracks relative to the number of stars we found between the 12 and 15$M_\odot$ tracks. Further, we compare this ratio to that predicted by the evolutionary models. Recall that the number of stars expected between masses $m_1$ and $m_2$ will be $$N_{m_1}^{m_2} = [m^\Gamma]_{m_1}^{m_2} \times \bar \tau$$ where $\Gamma$ is the slope of the initial mass function, taken here to be -1.35 (Salpeter 1955), and $\bar \tau$ is the average duration of the evolutionary phase for masses $m_1$ and $m_2$, shown in Table~\ref{tab:ages}. This equation assumes that the star formation rate has been relatively constant over the relevant time frame, which in this case is about 20 Myr, or the lifetime of a 12$M_\odot$ star. 

Table~\ref{tab:numbers} shows that both the models with no initial rotation (S0) and an initial rotation of 300 km s$^{-1}$ (S3) did a fairly good job predicting the number of 15 -- 25$M_\odot$ stars relative to the number of 12 -- 15$M_\odot$ stars. We observe a ratio of 1.6 to 2.0 and the S3 models predict a ratio of 1.6 while the S0 models predict 1.0. This is significantly better than what Drout et al.\ (2009) found in M31, where the number of 15 -- 25$M_\odot$ stars were a factor of 11 greater than predicted by the models, relative to the number of 12 -- 15$M_\odot$ stars. But, we observe no stars with masses above 25$M_\odot$ rather than the 3 to 6 predicted by the S0 or S3 models. This is similar to the case for M31 where the models predicted 110 -- 150 stars with masses above 25$M_\odot$, and none were observed. 

Now that we've compared the tracks in relative terms, we can compare the lifetimes in an absolute sense. Because we know the number of unevolved (OB-type) massive stars in the SMC and we now know the number of yellow supergiants, we can compute the expected lifetimes. According to a Schmidt survey of the bluest stars (Massey 2003), the number of unevolved SMC stars with masses greater than $20M_\odot$ is around 2600 (Massey 2009). The IMF-weighted H-burning lifetime is around 5 Myr, and if we assume a constant star formation rate, we would expect $5 \times 10^{-4}$ massive stars to be born per year. We can now make a comparison between the number of yellow supergiants greater than $20M_\odot$ observed (just one) vs.\ the number predicted by the Schmidt survey (2600). Recall that we were only able to observe 74\% of the 677 stars we selected. But, the Schmidt survey covered an area of the sky 73\% smaller than our surveyed area. So, these two percentages essentially cancel out. Therefore, we can estimate the actual ages of the yellow supergiant stage as 1/2600 $\times\ 5$ Myr. This works out to be around 1900 years, more than an order of magnitude lower than predicted by the evolutionary tracks (of order 0.1 Myr). We also tried this test with stars above $15M_\odot$. In this mass range there are around 4000 unevolved SMC stars and the duration is around 8 Myr. Since we found eight stars greater than $15M_\odot$, the age should be around 8/4000 $\times\ 8$ Myr, or 0.02 Myr. This \emph{still} corresponds to an order of magnitude lower than that predicted by the S3 models for the average duration of this stage. While the S0 models predict lower lifetimes, the relative durations predicted by the S3 models do a better job at matching our observed results. So, even though the model's relative durations are mostly correct, the time spent in the yellow supergiant stage is off by a factor of ten for both the S0 and S3 models.

This last test relies on our knowledge of the number of unevolved massive SMC stars. While the numbers are good approximations, they should be taken as lower limits due to the effects of crowding (Massey 2003). Still, we estimate that they are probably good to a factor of a few. Thus, while our current prediction is that the models are off by a factor of ten, the actual error is probably somewhat lower due to uncertainties in the number of unevolved SMC stars.

\section{Summary and Conclusions}
\label{C}
After selecting 677 potential SMC supergiants, we observed 498. We identify 176 stars as candidate SMC yellow supergiants and 16 stars as possible SMC yellow supergiants while the rest are categorized as foreground stars. Our literature search confirmed that we identified nearly all of the SMC yellow supergiants down to $12M_\odot$. Additionally, the Besan\c{c}on models suggest a Milky Way halo contamination of less than 4\% of our 176 SMC supergiants.

\begin{figure}
\epsscale{1}
\plotone{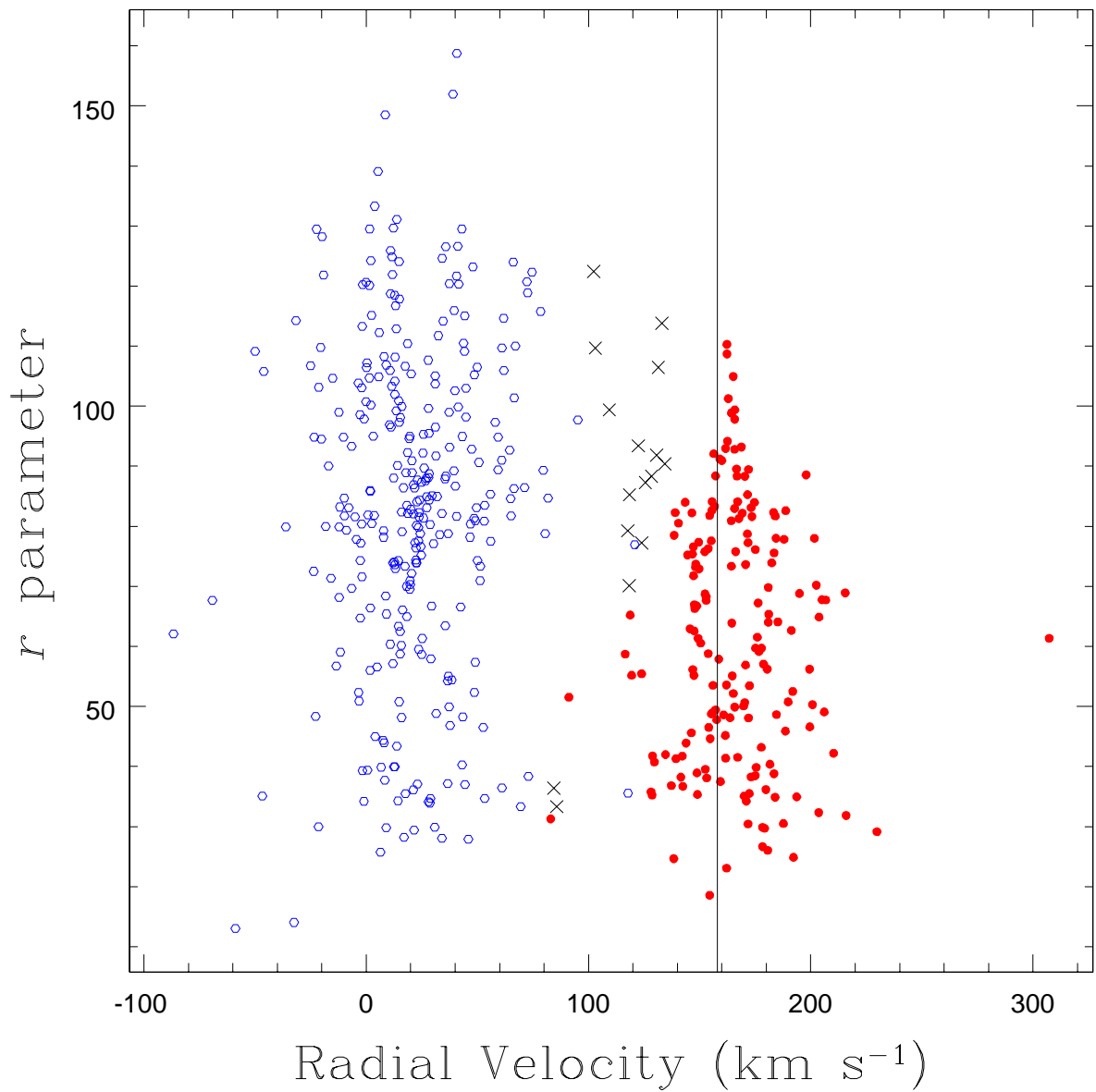}
\caption{\label{fig:radVelRplotMem} Radial velocity membership determination. In this figure, the red filled circles represent category 1 stars, the black $\times$s represent category 2 stars and the blue unfilled hexagons represent category 3 stars. Again, the vertical line is at 158 km s$^{-1}$ representing the average radial velocity of the SMC. The outlier with a radial velocity near 300 km s$^{-1}$ could either be a binary or a runaway. However, we are confident that this radial velocity measurement is correct.}
\end{figure}

We also note that, as shown in Figure~\ref{fig:radVelRplotMem}, the median radial velocity of our category 1 SMC supergiants doesn't quite fall on the line denoting the published systemic radial velocity of the SMC (158~km~s$^{-1}$) (Richter et al.\ 1987). Instead, our calculated median is 166.0 km s$^{-1}$ with a spread (standard deviation) of 24.3 km s$^{-1}$. Evans \& Howarth (2008) found similar radial velocity results when studying the kinematics of massive stars in the SMC. They found the average radial velocity of F and G type stars to be 160.8 $\pm$ 0.5 km s$^{-1}$ where $\sigma =$ 35.1 km s$^{-1}$ and the average radial velocity of all O,B,A,F and G type stars in their sample to be 172.0 $\pm$ 0.2 where $\sigma =$ 33.6 km s$^{-1}$. The large sigmas of both studies reflect the true range of the SMC's radial velocity due to the complicated kinematics of the SMC as revealed by the HI maps of Stanimirovi\'{c} et al.\ (2004). 

We also placed the supergiants on the HRD and compared our results to the Geneva evolutionary tracks. In terms of the stars' locations relative to the tracks, we found good agreement. In terms of the relative number of different mass stars, we found that the ratio for 15 -- 20$M_\odot$ stars to 12 -- 15$M_\odot$ stars agreed almost perfectly. However for stars greater than 20$M_\odot$, this comparison quickly begins to fail. It is interesting to mention here that the apparent lack of yellow supergiants in the upper HRD occurs at the position where the ``yellow evolutionary void" defined by Nieuwenhuijzen and de Jager (1995) happens to be. These authors find that this void corresponds to an instability region for blue-ward evolving stars. It might be that stars in this region undergo strong outbursts which will make them evolve rapidly away from it. We shall study this point in a forthcoming paper. Lastly, in terms of the actual timescales that these stars stay in the yellow supergiant phase, these numbers appear to be off by factors of ten over all mass ranges.

The timescales for the F and G supergiant phases are similarly overestimated compared to the data for both M31 and the SMC, suggesting that wherever the problem lies, it is not purely with the treatment of mass-loss on the main-sequence. Where then does the problem lie? We considered at first that the problem could reside with the assumed mass-loss during the red supergiant phase, which would affect the time-scales in the cases that the models predict the stars to evolve back to the blue. But, an examination of the models shows that the vast majority of the predicted lifetimes of the yellow supergiant stage is spent in the evolution from the blue side to the red side. Therefore, uncertainties during the RSG cannot be responsible. We have now experimented with newer (unpublished) Geneva models and find better agreement between the predicted and observed values. Pinpointing the major effect affecting the yellow supergiant lifetimes will require more trial calculations. While these theoretical tests are being conducted, we plan to extend the observational work to the LMC, where the sample size is larger and where the metallicity is around two times greater than that of the SMC.

\acknowledgements
We would like to acknowledge the generous allocation of observing time by the NOAO Time Allocation Committee, and the excellent support we received while observing with Hydra on the Blanco, particularly by Ricardo Venegas and Mauricio Rojas Gonzalez.  This work was partially supported by the National Science Foundation through AST-0604569. We thank Dr.\ Deidre Hunter and Dr.\ Steve Slivan for their critical readings of the manuscript as well as the anonymous referee for useful comments and suggestions which improved the manuscript.

\begin{deluxetable}{c c c r c r c c r r c c c c l}
\tabletypesize{\tiny}
\rotate
\tablecaption{\label{tab:allobs} Properties of Observed Targets\tablenotemark{*}}
\tablewidth{0pt}
\tablehead{
\colhead{2MASS}
&\colhead{$\alpha$(J2000)}
&\colhead{$\delta$(J2000)}
&\colhead{$K$}
&\colhead{$J-K$}
&\colhead{$V$}
&\colhead{$B-V$}
&\colhead{Ref.\tablenotemark{b}}
&\colhead{Vel}
&\colhead{$r$\tablenotemark{c}}
&\colhead{Category\tablenotemark{d}}
&\colhead{Hydra}
&\colhead{Literature}
&\colhead{Ref.\tablenotemark{e}}
&\colhead{Comments\tablenotemark{f}} \\
& & & & & & & & 
\colhead{(km s$^{-1}$)}
& & &
\colhead{Class}
&\colhead{Class}
}
\startdata
J00342570-7322334 & 00 34 25.69 & -73 22 33.5 & 11.60 & 0.25 & 12.79 & \nodata & 3 &   2.0 & 100.2 & 3 & \nodata   & \nodata    & \nodata &  \nodata                          \\
J00342883-7328192 & 00 34 28.82 & -73 28 19.3 & 10.72 & 0.59 & 13.03 & 0.97    & 4 & 103.1 & 109.7 & 2 & \nodata   & \nodata    & \nodata &  \nodata                          \\
J00344094-7319366 & 00 34 40.94 & -73 19 36.7 &  9.18 & 0.69 & 11.73 & \nodata & 3 &  50.0 & 106.5 & 3 & \nodata   & \nodata    & \nodata &  \nodata                          \\
J00345529-7213044 & 00 34 55.30 & -72 13 04.4 & 11.58 & 0.31 & 12.93 & \nodata & 3 &   1.7 &  56.0 & 3 & \nodata   & \nodata    & \nodata &  \nodata                          \\
J00345992-7339156 & 00 34 59.91 & -73 39 15.8 & 10.66 & 0.33 & 12.13 & \nodata & 3 &  21.9 &  77.3 & 3 & \nodata   & F8         & 6       &  Flo 72                           \\
J00350326-7332586 & 00 35 03.25 & -73 32 58.8 & 10.77 & 0.80 & 13.80 & 1.54    & 4 & 146.8 &  75.4 & 1 & \nodata   & \nodata    & \nodata &  SkKM 2                           \\
J00351576-7340108 & 00 35 15.75 & -73 40 10.9 & 10.25 & 0.47 & 12.13 & 0.85    & 4 &   2.4 & 115.1 & 3 & \nodata   & \nodata    & \nodata &  \nodata                          \\
J00353473-7341121 & 00 35 34.74 & -73 41 12.3 & 11.04 & 0.37 & 12.55 & \nodata & 3 &  -4.4 &  77.8 & 3 & \nodata   & \nodata    & \nodata &  \nodata                          \\
J00353564-7205134 & 00 35 35.64 & -72 05 13.5 & 11.28 & 0.34 & 12.70 & \nodata & 3 &  14.5 &  63.3 & 3 & \nodata   & \nodata    & \nodata &  \nodata                          \\
\enddata
\tablenotetext{*}{The full version of this table can be found online.}
\tablenotetext{a}{These stars were imaged twice and their results were averaged. The average difference in Vel$_{\rm obs}$ was 2.2 km s$^{-1}$ and the average difference in the $r$ parameter was 13.7.}
\tablenotetext{b}{References for $B-V$ colors and $V$ magnitudes. 1 = Massey 2002; 2 = Mermilliod 1997; 3 = ASAS-3; 4 = Zaritsky 2002; 5 = (Zaritsky B) - (ASAS-3 V)}
\tablenotetext{c}{Tondry \& Davis 1979 $r$ parameter.}
\tablenotetext{d}{Category: 1 = SMC supergiant; 2 = possible SMC supergiant; 3 = foreground dwarf}
\tablenotetext{e}{References for spectral classifications. 1 = Cannon \& Pickering 1918; 2 = Massey \& Olsen 2003; 3 = Evans et al.\ 2004; 4 = Sanduleak 1969; 5 = L\"{u} 1971; 6 = Florsch 1972; 7 = Feast 1974; 8 = Azzopardi et al.\ 1975; 9 = Houk \& Cowley 1975; 10 = Ardeberg \& Maurice 1977; 11 = Azzopardi \& Vigneau 1979; 12 = Trundle \& Lennon 2005; 13 = Humphreys 1983; 14 = Levesque et al.\ 2006; 15 = Wallerstein 1984; 16 = Bouchet et al.\ 1985; 17 = Carney et al.\ 1985; 18 = Massey et al.\ 2009; 19 = Sanduleak 1989; 20 = Humphreys et al.\ 1991; 21 = Lennon 1997}
\tablenotetext{f}{Comments include: alternate names for the star, comments on the blue spectra and comments based on the literature search.}
\end{deluxetable}

\begin{deluxetable}{l c c c r}
\tablecaption{\label{tab:derived} Derived Properties of SMC Yellow Supergiants\tablenotemark{*}}
\tablewidth{0pt}
\tablehead{
\colhead{2MASS}
&\colhead{Category}
&\colhead{$T_{\rm eff}$}
&\colhead{$\log{(L/L_{\sun})}$}
&\colhead{Equivalent Width} \\
& & &
&\colhead{of OI $\lambda$7774 (\AA)\tablenotemark{b}}
}
\startdata
J00490296-7321409 &  1 & 4.073 & 5.527 & 0.6 \\
J00503839-7328182 &  1 & 4.044 & 5.435 & $<0.2$      \\
J00530489-7238000 &  1 & 3.973 & 5.353 & 0.8 \\
J01044935-7206218 &  1 & 4.044 & 5.307 & 0.2 \\
J01055631-7219448 &  1 & 4.058 & 5.302 & 0.3 \\
J01024960-7210145 &  1 & 3.937 & 5.279 & 1.2 \\
J01240629-7314454\tablenotemark{a} &  1 & 3.693 & 5.266 & $<0.2$      \\
J00411604-7232167\tablenotemark{a} &  1 & 3.629 & 5.253 & $<0.2$      \\
J00525121-7306535 &  1 & 3.901 & 5.223 & 1.6 \\
J00530894-7229386 &  1 & 3.604 & 5.219 & $<0.2$      \\
\enddata
\tablenotetext{*}{The full version of this table can be found online.}
\tablenotetext{a}{This star didn't have a known $B-V$ color, so instead we used the star's $J-K$ color to derive the $T_{\rm eff}$ and $\log{(L/L_{\sun})}$.}
\tablenotetext{b}{We found we couldn't measure OI $\lambda$7774 if it had an equivalent width of less than 0.2 \AA.}
\end{deluxetable}

\begin{deluxetable}{c c c r c c c l c l l}
\tablecaption{\label{tab:knownSG} SMC Supergiants Not Observed}
\tabletypesize{\footnotesize}
\rotate
\tablewidth{0pt}
\tablehead{
\colhead{2MASS}
&\colhead{$\alpha_{\rm 2000}$}
&\colhead{$\delta_{\rm 2000}$}
&\colhead{$K$}
&\colhead{$J-K$}
&\colhead{$V$}
&\colhead{Ref.\tablenotemark{b}}
&\colhead{Lit.}
&\colhead{Ref.\tablenotemark{c}}
&\colhead{Alt. Names}
&\colhead{Comment\tablenotemark{d}}}
\startdata
J00522260-7239509 & 00 52 22.60 & -72 39 51.1 & 9.56 & 0.43 & 11.31 & 2 & F5Ia & 5 & AzV 121 & double star \\
J01291727-7243202 & 01 29 17.27 & -72 43 20.2 & 10.19 & 0.34 & 11.56 & 3 & F0Iae & 5 & Sk 181 & outside selected region \\
J01045507-7202364 & 01 04 55.08 & -72 02 36.4 & 9.71 & 0.33 & 11.85 & 2 & F5Ia & 5 & Sk 118; AzV 369 & double star \\
J00530194-7138375\tablenotemark{a} & 00 53 01.93 & -71 38 37.6 & 10.46 & 0.34 & 11.87 & 2 & F3Iab & 4 & Sk 55; AzV 140 \\
J01041548-7245202\tablenotemark{a} & 01 04 15.48 & -72 45 20.3 & 9.91 & 0.56 & 11.91 & 1 & G2Ib & 2 & SkKM 269 \\
J00524559-7304242\tablenotemark{a} & 00 52 45.61 & -73 04 24.3 & 12.33 & 0.09 & 12.77 & 1 & F2I & 1 & AzV 134 \\
J00525682-7155032\tablenotemark{a} & 00 52 56.81 & -71 55 03.2 & 10.54 & 0.54 & 12.78 & 2 & G0Ib & 2 & HV 11157 \\
J00510935-7248358\tablenotemark{a} & 00 51 09.35 & -72 48 35.9 & 12.43 & 0.13 & 12.95 & 1 & F2I & 1 & AzV 88 \\
J01061522-7235284 & 01 06 15.22 & -72 35 28.5 & 12.33 & 0.24 & 13.20 & 1 & F2I & 1 & AzV 401 & $K$ magnitude too faint \\
J01030066-7230121 & 01 03 00.66 & -72 30 12.2 & 12.13 & 0.26 & 13.24 & 1 & F2I & 1 & AzV 323 & $K$ magnitude too faint \\
J00562532-7228182\tablenotemark{a} & 00 56 25.32 & -72 28 18.2 & 10.03 & 0.89 & 13.32 & 1 & G5Ib & 3 & SkKM 153 \\
J00580171-7219356 & 00 58 01.72 & -72 19 35.5 & 12.07 & 0.29 & 13.41 & 1 & F5I & 1 & AzV 198 & $K$ magnitude too faint \\
J00542773-7215580 & 00 54 27.76 & -72 15 58.0 & 12.54 & 0.20 & 13.56 & 1 & F2I & 1 & AzV 159 & $K$ magnitude too faint \\
J00514592-7235533 & 00 51 45.90 & -72 35 53.3 & 12.36 & 0.34 & 13.63 & 1 & F5I & 1 & AzV 107 & poor color quality codes \\
J00504058-7250030 & 00 50 40.58 & -72 50 03.0 & 12.91 & 0.17 & 13.71 & 1 & F2I & 1 & AzV 79 & $K$ magnitude too faint \\
J01072633-7221511 & 01 07 26.32 & -72 21 51.4 & 12.63 & 0.28 & 13.93 & 1 & F5:I & 1 & AzV 417 & $K$ magnitude too faint \\
J00501412-7243404 & 00 50 14.11 & -72 43 40.4 & 13.15 & 0.24 & 14.03 & 1 & F5I & 6 & AzV 67a & $K$ magnitude too faint \\
\enddata
\tablenotetext{a}{Stars we selected but were not observed.}
\tablenotetext{b}{References for $V$ magnitudes. 1 = Massey 2002; 2 = Mermilliod 1997; 3 = ASAS-3}
\tablenotetext{c}{References for spectral classifications. 1 = Azzopardi et al.\ 1975; 2 = Wallerstein 1984; 3 = Feast 1979; 4 = Ardeberg \& Maurice 1977; 5 = Humphreys 1983; 6 = Azzopardi \& Vigneau 1979.}
\tablenotetext{d}{Reason why this star wasn't in our original sample.}
\end{deluxetable}

\begin{deluxetable}{l r r}
\tablecaption{\label{tab:ages}SMC Theoretical Yellow Supergiant Duration\tablenotemark{a}}
\tablewidth{0pt}
\tablehead{
\colhead{$M_\odot$} 
&\colhead{S3\tablenotemark{b}} 
&\colhead{S0\tablenotemark{b}} \\
& \colhead{(years)}
&\colhead{(years)}
}
\startdata
60 & 24,600 & 56,500 \\
40 & 324,300 & 32,700 \\
25 & 118,900 & 15,000 \\
20 & 71,700 & 16,500 \\
15 & 206,600 & 60,400 \\
12 & 33,200 & 30,700 \\
9   & 490,400 &  71,500 \\
\enddata
\tablenotetext{a}{For the purposes of this calculation, yellow supergiants are defined as having an $T_{\rm eff}$ between 4800~K and 7500~K.}
\tablenotetext{b}{S3 has an initial rotation of 300 km s$^{-1}$, and S0 has no initial rotation. Ages were determined using models from Maeder \& Meynet (2001).}
\end{deluxetable}

\begin{deluxetable}{l c c c c c c}
\tabletypesize{\small}
\tablecaption{\label{tab:numbers} Number of Yellow Supergiants: Observed vs. Modeled\tablenotemark{a}}
\tablewidth{0pt}
\tablehead{
\colhead{Mass}
&\colhead{\#}
&\colhead{\#}
&\multicolumn{4}{c}{Ratio relative to 12-15$M_\odot$}  \\ \cline{4-7}
\colhead{Range}
&\colhead{All}
&\colhead{Certain}
&\colhead{All}
&\colhead{Certain}
&\colhead{S3}
&\colhead{S0}
}
\startdata
12-15$M_\odot$   & 5  & 3 & 1.0 & 1.0 & 1.0 & 1.0 \\
15-25$M_\odot$   & 8  & 6 & 1.6 & 2.0 & 1.6 & 1.0 \\
25-40$M_\odot$   & 0  & 0 & 0.0 & 0.0 & 1.2 & 0.5 \\
40-60$M_\odot$   & 0  & 0 & 0.0 & 0.0 & 0.5 & 1.0 \\
\enddata
\tablenotetext{a}{Models were run with $z = 0.004$. S0 models were run with an initial rotation of 0 km s$^{-1}$ while S3 models were run with an initial rotation of 300 km s$^{-1}$.}
\end{deluxetable}

\end{document}